\documentclass[pdflatex,sn-mathphys-num]{sn-jnl}% Math and Physical Sciences Numbered Reference Style
\usepackage{graphicx}%
\usepackage{multirow}%
\usepackage{amsmath,amssymb,amsfonts}%
\usepackage{amsthm}%
\usepackage{mathrsfs}%
\usepackage[title]{appendix}%
\usepackage{xcolor}%
\usepackage{textcomp}%
\usepackage{manyfoot}%
\usepackage{booktabs}%
\usepackage{algorithm}%
\usepackage{algorithmicx}%
\usepackage{algpseudocode}%
\usepackage{listings}%
\usepackage{caption}
\usepackage{subcaption}
\usepackage{braket}

\theoremstyle{thmstyleone}%
\theoremstyle{thmstyletwo}%

\theoremstyle{thmstylethree}%

\begin{document}

\title[A machine learning based approach to the identification of spectral densities in quantum open systems]{A machine learning based approach to the identification of spectral densities in quantum open systems}


\author[1]{\fnm{Jessica} \sur{Barr}}\email{jbarr24@qub.ac.uk}

\author[2]{\fnm{Shreyasi} \sur{Mukherjee}}\email{shreyasi.mukherjee@dfa.unict.it}

\author[3]{\fnm{Alessandro} \sur{Ferraro}}\email{alessandro.ferraro@unimi.it}

\author*[4,1]{\fnm{Mauro} \sur{Paternostro}}\email{mauro.paternostro@unipa.it}

\author*[1]{\fnm{Giorgio} \sur{Zicari}}\email{G.Zicari@qub.ac.uk}

\affil*[1]{\orgdiv{Centre for Quantum Materials and Technologies, School of Mathematics and Physics}, \orgname{Queen's University Belfast}, \orgaddress{\street{University Road}, \city{Belfast}, \postcode{BT7 1NN}, \country{United Kingdom}}}

\affil[2]{\orgdiv{Dipartimento di Fisica e Astronomia  ``Ettore Majorana''}, \orgname{Universit\`a di Catania}, \orgaddress{\street{Via S. Sofia 64}, \city{Catania}, \postcode{95123}, \country{Italy}}}

\affil[3]{\orgdiv{Dipartimento di Fisica ``Aldo Pontremoli''}, \orgname{Universit\`{a} degli Studi di Milano}, \orgaddress{\street{Via Celoria 16}, \city{Milano}, \postcode{I-20133}, \country{Italy}}}

\affil[4]{\orgdiv{Quantum Theory group, Dipartimento di Fisica e Chimica ``Emilio Segr\`e''}, \orgname{Universit\`a degli Studi di Palermo}, \orgaddress{\street{via Archirafi 36}, \city{Palermo}, \postcode{I-90123}, \country{Italy}}}


\abstract{We present a machine learning-based approach for characterising the environment that affects the dynamics of an open quantum system. We focus on the case of an exactly solvable spin-boson model, where the system-environment interaction, whose strength is encoded in the spectral density, induces pure dephasing. By using artificial neural networks trained on the Fourier-transformed time evolution of some observables of the system, we perform both classification --distinguishing sub-Ohmic, Ohmic, and super-Ohmic spectral densities -- and regression --- thus estimating key parameters of the spectral density function, when the latter is expressed through a power law. Our results demonstrate high classification accuracy and robust parameter estimation, highlighting the potential of machine learning as a powerful tool for probing environmental features in quantum systems and advancing quantum noise spectroscopy.}

%%================================%%
%% Sample for structured abstract %%
%%================================%%




\maketitle

\section{Introduction}\label{sec1}

Quantum systems are inherently open. Conceptually motivated by the difficulty in isolating a quantum system from various sources of external noise, one often regards a quantum system as coupled to a large external environment that induces dissipation and/or decoherence~\cite{Breuer-Petruccione,Vacchini:2024}. In standard scenarios, the coupling between the system and the environment allows for a neat separation of the relevant timescales, leading to a Markovian description of the dynamics. However, this is not the case in a variety of scenarios, where strong-coupling and non-Markovian effects naturally emerge~\cite{DeVega:2017}.
Knowledge of the environment is indeed essential to describe open system dynamics. On one hand, it allows to limit the detrimental effects of dissipation and decoherence, which constitute a major obstacle in the realisation of quantum technologies. On the other hand, it might be used to exploit or engineer dissipation for optimising quantum control strategies~\cite{Poyatos:1996,Verstraete:2009,Harrington:2022}.

In the open quantum system formalism,  information about the system-environment coupling is encoded in the so-called  spectral density (SD), which enters the expression of the environmental correlation function, and thus ultimately affects the temporal behavior of the system's observables. 

Building on the success of recent approaches leveraging machine learning (ML) to harness some aspects of noise affecting the dynamics of an open quantum system~\cite{garau2019machine2,zhang2019spin,Youssry2020,PRXQuantum.2.010316,Palmieri:2021,papivc2022neural2,martina2022learning,martina2023machine,martina2023deep,mukherjee2024noise}, we aim to show that ML can be fruitfully deployed to characterize the salient features of the SD of an environment. Specifically, we focus on the case of a class of spin-boson (SB) models where the coupling between system and environment induces pure dephasing dynamics.

Some of us have shown in Ref.~\cite{barr2024spectral} that, for this class of SB models, an artificial neural network (NN) can be used to process the time-series of a system observable and classify %the SD. More specifically, the focus is restricted to 
power-law SDs, %thus three classes were considered, i.e. 
as Ohmic, sub-Ohmic, or super-Ohmic. While successful in achieving accurate classification, this approach did not provide information on the parameters entering the functional form of the SD. Here, we go beyond the context set by Ref.~\cite{barr2024spectral} and perform regression on the relevant parameters of a SD function, thus providing a more rounded characterisation of the environmental features.

The remainder of the paper is organised as follows. Sec.~\ref{sec:setting} introduces the general theoretical setting, discussing the general features of spin-boson models where the bosonic bath is a large bosonic reservoir, with a focus on an exactly solvable model describing pure dephasing dynamics. Sec.~\ref{sec:methods} outlines the methodology, including the machine learning framework employed for classification and regression tasks. Sec.~\ref{sec:results} presents the analysis and results of our numerical experiments. Finally, we conclude with a discussion of our findings and potential directions for future studies.

\section{General setting}\label{sec:setting}

Let us consider the case of a two-level system coupled to an external bosonic bath comprising a large number of  bosonic degrees of freedom. The full system-environment Hamiltonian reads
%\begin{equation}
 %   \label{eq:SB_hamiltonian}
  $  H = H_S + H_B + H_I$,
%\end{equation}
where % which we assume to be given by
\begin{equation}
    \label{eq:H_sys}
    H_S = \frac{\omega_0}{2} \sigma_z 
\end{equation}
is the system Hamiltonian with  $\omega_0$ the bare frequency of the two-level system, and $\sigma_z$ is the $z$ Pauli operator. The environment is modelled as a discrete set of independent harmonic oscillators, i.e.
\begin{equation}
    \label{eq:B_hamiltonian}
    H_B = \sum_{k} \omega_k b_k^\dagger b_k,
\end{equation}
where $b_k, b_k^\dagger$ are the annihilation and creation operators associated with the $k$-th bosonic mode. To keep the discussion general, we assume that the system-environment coupling has the form
\begin{equation}
    \label{eq:H_int}
    H_I = X \otimes B \, ,
\end{equation}
where $X$ is a system observable and $B$ is a bath one, which we assume to be a linear combination of bosonic modes. With this in mind, we set
\begin{equation}
    \label{eq:bath_operator}
    B = \sum_k \left ( g_k b_k^\dagger + g_k^* b_k \right ) \, ,
\end{equation}
where $g_k$ is the rate of interaction between the system and the $k$-th environmental mode through the observable $X$. Such coupling constants control the overall system-environment interaction, as they appear in the formal definition of the SD, i.e.
\begin{equation}
    \label{eq:SD_def}
    J(\omega) = \sum_k |g_k|^2 \delta (\omega - \omega_k) \, .
\end{equation}
For bosonic environments, the two-time bath correlation function defined as $C_\beta(t) = \langle B(t) B(0)  \rangle_B$ depends on the SD, where $B(t)$ is the bath operator in the interaction picture with respect to the free Hamiltonian $H_0 = H_S + H_B$. In fact, one can show that, in the limit where the environmental modes form a continuum, the correlation function reads 
\begin{equation}
    \label{eq:correlation_function}
    C_\beta(t) = \int\limits_0^\infty d\omega J(\omega) \left [ \coth{\left ( \frac{\beta \omega}{2} \right )} \cos{\omega t} - i \sin{\omega t}\right],
\end{equation}
provided that the environment is prepared in a Gibbs state $\rho_B = e^{-\beta H_B}/\mathcal{Z}_B$ at inverse temperature $\beta$ (here $\mathcal{Z}_B \equiv \operatorname{tr}_B [e^{- \beta H_B}]$ is the reservoir partition function)~\cite{Vacchini:2024,barr2024spectral}. 

Let us focus on the case where the system operator in Eq.~(\ref{eq:H_int}) is given by $X= \sigma_z$, so that %the interaction Hamiltonian commutes with the total Hamiltonian as 
$[H, \sigma_z] =0$. In particular, the interaction term commutes with the free system Hamiltonian, which is therefore a constant of motion, implying time independence of the populations. Starting from an initial product state for the system-environment composite system such as $\rho_{SB}(0) = \rho(0) \otimes \rho_B$, this symmetry allows for an exact elimination of the environmental degrees of freedom~\cite{Breuer-Petruccione,Palma:96}. %, where $\rho_B = e^{-\beta H_B}/\mathcal{Z}_B$ is a thermal Gibbs state, and. 
Using the basis $\{ \ket{0}, \ket{1}\}$ of eigenstates of $\sigma_z$, the evolved state of the system can be written as~\cite{Vacchini:2024,barr2024spectral}
\begin{equation} \label{eq:Exactlysolvablerhot}
    \rho(t) = \begin{pmatrix}
    \rho_{00}^0 & \rho_{01}^0  e^{- \Gamma (t)} \\
    \rho_{01}^{0*} e^{- \Gamma (t)} & 1-\rho_{00}^0
    \end{pmatrix} \, ,
\end{equation}
where $\rho_{ij}^0=\bra{i}\rho(0)\ket{j}~(i,j=0,1)$ are the entries of the initial density matrix $\rho(0)$. The thermal bath influences coherence through the \emph{decoherence function}
\begin{equation}\label{eq:decoherencefunction}
    \Gamma (t) = 4 \int\limits_0^{\infty} \textrm{d} \omega J (\omega) \coth \left( \frac{\beta \omega}{2} \right) \frac{1 - \cos (\omega t)}{\omega^2} \, .
\end{equation}
Depending on the form of the SD, the function $\Gamma(t)$ can be either positive or negative over specific intervals of time: in the former case, coherences are exponentially damped over time, while in the latter the system is able to \textit{re}-cohere, a property that can be seen as a signature of non-Markovianity~\cite{Addis:2014,Guarnieri:2014}.

%Let us consider the case where the system couples to the environment through the first Pauli operator, i.e. $X= \sigma_x/2$. Since explicit symmetries are not present in this case, one cannot discard the environmental degrees of freedom in an exact manner.  However, starting from an uncorrelated initial state, one can derive a master equation, effectively capturing the influence of the interaction with the bosonic environment up to the second order in the coupling constant. Qualitatively, along with coherences, one should expect that the populations evolve over time due to the interaction with the thermal bath, possibly reproducing a thermalisation process. In this case, we can decompose the correlation function as
%\begin{equation}
    %\label{eq:correlation_function_ampl}
   % C_\beta(t) = \nu(t) + i \mu(t) \, ,
%\end{equation}
%where $\nu(t) = \textrm{Re} [ C_\beta(t)]$ and $ \mu(t) = \textrm{Im} [ C_\beta(t)]$ are the noise and dissipation kernels, respectively, which are related by the corresponding fluctuation-dissipation relation~\cite{Pottier:2010,HU:1992}.

\section{Methods}
\label{sec:methods}

\begin{figure}[t!]
     \centering
     \begin{subfigure}[b]{0.45\textwidth}
         \centering
         \includegraphics[width=\textwidth]{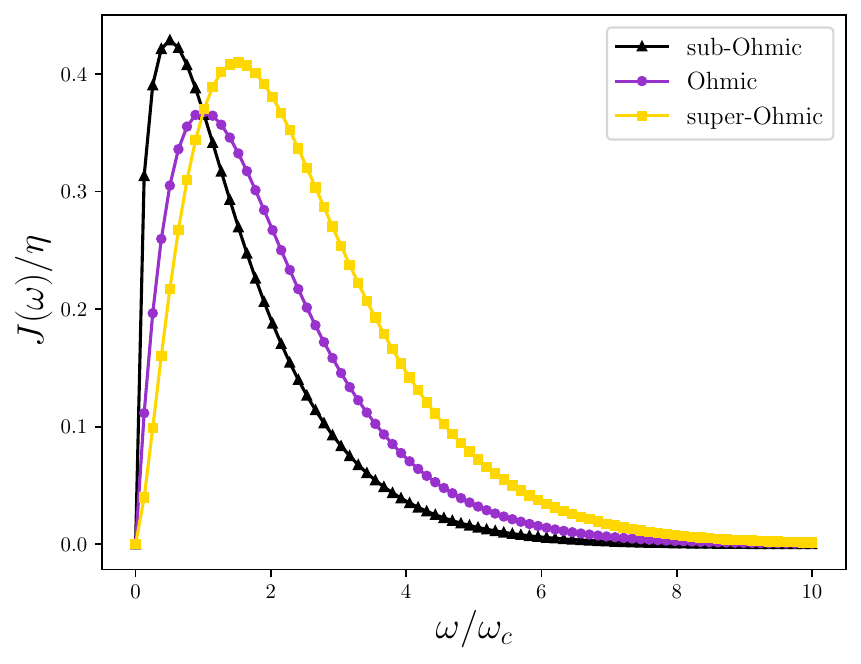}
         \caption{Spectral density}
         \label{fig:SD}
     \end{subfigure}
     \hfill
     \begin{subfigure}[b]{0.47\textwidth}
         \centering
         \includegraphics[width=\textwidth]{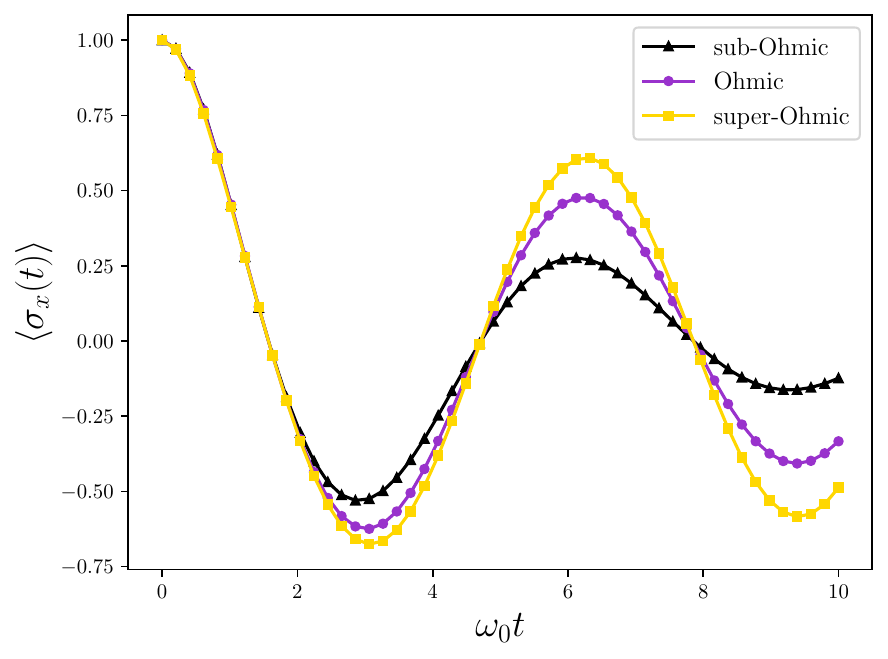}
         \caption{Pure dephasing}
         \label{fig:deph_curves}
     \end{subfigure}
        \caption{Panel~(a): examples of spectral densities $J(\omega)$ in the form of Eq.~(\ref{eq:SD_general_s}). The curves corresponding to the sub-Ohmic, Ohmic, and super-Ohmic case are generated choosing $s=0.5,1,1.5$, respectively. Panel~(b): instance of the behavior of $\langle \sigma_x(t) \rangle$ under the pure dephasing model described in Sec.~\ref{sec:setting} and for different choices of SD [$\eta=0.1$, $\omega_c=1$ and the same values of $s$ as in panel (a)]. We consider a zero-temperature bath and start from the initial state $\rho(0)=| + \rangle \! \langle + |$, with $\ket{+} = (\ket{0} + \ket{1})/\sqrt{2}$. }
        \label{fig:three graphs}
\end{figure}

 Given a SB model of the types described in Sec. \ref{sec:setting}, we focus on a specific class of SDs, described by a power law of the form~\cite{Weiss:2012,DeVega:2017}
 \begin{equation}
     \label{eq:SD_general_s}
     J(\omega) = \eta \, \omega_c^{1-s} \omega^s \, e^{-\frac{\omega}{\omega_c}},
 \end{equation}
 where $s$ is the so-called \emph{Ohmicity parameter}, $\eta$ is the coupling constant between the system and the environment, and $\omega_c$ is a cut-off frequency. Note that we choose an exponential cut-off ensuring that $J(\omega) \to 0 $ as $\omega \to \infty$.
 The Ohmicity parameter is a crucial quantity, as it ultimately controls the system-environment interaction. Depending on the value of $s$, the SD is said to be \emph{Ohmic} ($s=1$), \emph{sub-Ohmic} ($s<1$), or \emph{super-Ohmic} ($s>1$). Some instances of this type of SDs are shown in Fig.~\ref{fig:SD}.
 For a given functional form of the SD, one can in principle solve the dynamics of the system, as well as track the evolution of any of its observables. The expectation value of a generic observable $O$ is formally obtained as
 \begin{equation}
     \label{eq:time_obs}
     \langle O(t) \rangle \equiv \operatorname{Tr}_{SB} \left ( O e^{- i H t} \,  \rho_{SB}(0) \, e^{i H t}\right ),
 \end{equation}
 where, as before, %$H$ is the full system-environment Hamiltonian, while
 the global initial state is factorised as $\rho_{SB}(0) = \rho(0) \otimes \rho_B$. %, with $\rho(0)$ and $\rho_B$ being the initial system and environmental states, respectively. 
 It can be shown that the expectation value $\langle O(t) \rangle$  depends on the choice of $J(\omega)$~\cite{Mascherpa:2017}. 

 As we consider a two-level system, we choose as relevant observables the three Pauli matrices $\sigma_j$ (with $j=x,y,z$). For instance, for the pure-dephasing model described in Sec.~\ref{sec:setting}, $\langle \sigma_z \rangle$ depends trivially on $J(\omega)$, as $\sigma_z$ is a constant of motion, whereas one can show that $\langle \sigma_x \rangle$ and $\langle \sigma_y \rangle$ depend on the SD through the decoherence function $\Gamma(t)$. Explicitly, one has
 \begin{equation}
     \label{eq:deph_obs}
      \langle \sigma_x (t)\rangle = {\rm Re}\{v(t)\}%2 \, {\rm Re} \{ \rho_{01}^0\} \, e^{-\Gamma(t)} 
      \quad \text{and} \quad \langle \sigma_y (t)\rangle = - {\rm Im}\{v(t)\}.
      %2 \, {\rm Im} \{ \rho_{01}^0\} \, e^{-\Gamma(t)} \,.
 \end{equation}
 with $v(t)=2\rho^0_{01}e^{-\Gamma(t)}$.
In Fig.~\ref{fig:deph_curves}, we plot $\langle \sigma_x (t) \rangle$ as a function of time corresponding to the three different classes: sub-Ohmic, Ohmic, super-Ohmic.

 The next step is to resort to an artificial NN to process data and perform both classification and regression tasks on them. The dataset is generated by solving the system dynamics. Specifically, we consider the expectation value of a given system observable $\langle O (t)\rangle$ sampled at $N$ points $t_n$ ($n=0, \ldots, N-1$). However, instead of using the time series $\{\langle O (t_n) \rangle\}_{n=0,\ldots, N-1}$, we first pre-process the input data by Fourier-decomposing the signal, thus achieving the Fourier coefficients% are given by
 \begin{equation}
     \label{eq:Fourier_coeff}
     X_{k} = \sum_{n=0}^{N-1} \langle O (t_n) \rangle e^{-i\frac{2 \pi k n}{N}} \, ,
 \end{equation}
 where $k = 0, \ldots, N-1$. The original signal is obtained by formally inverting Eq.~(\ref{eq:Fourier_coeff}). 
In general, the Fourier coefficients $X_{k}$ are complex numbers, so to faithfully reconstruct the original signal we would need both the real and the imaginary part.

\begin{figure}[t!]
\centering
\includegraphics[width=0.9\textwidth]{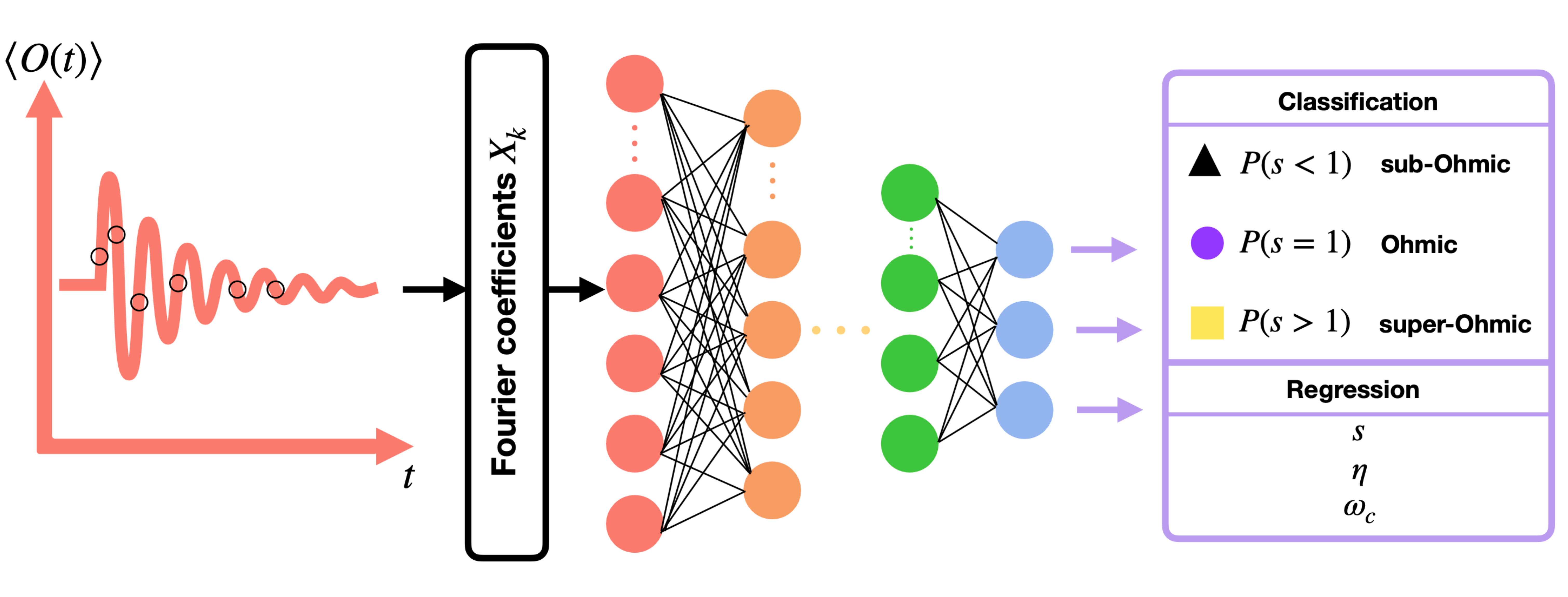}
\caption{Sketch of the setup. Given the time evolution of a system observable, denoted as $\langle O(t) \rangle$, we compute the corresponding Fourier coefficients
${X_k}$, which are fed into a NN to infer properties of the SD of the form given by Eq.~(\ref{eq:SD_general_s}). For the classification task, the output layer contains three neurons, each representing the probability that the SD is sub-Ohmic ($s<1$), Ohmic ($s=1$), or super-Ohmic ($s>1$). For the regression task, the output layer has a number of neurons matching the number of parameter we want to determine, i.e. $s, \eta, \omega_c$.}
\label{fig:sketch}
\end{figure}

The dataset containing all the Fourier coefficients is used as an input for the artificial NN~\cite{MacKay:2003,Marquardt:2021}. The latter consists of a large set of computational units, known as artificial neurons, arranged in a series of layers. The first layer, which only receives the input data without manipulating them, is known as \emph{input layer}. The data are then passed to a finite number of layers -- known as \emph{hidden layers} -- where each neuron receives a set of inputs $\{x_i\}$ and computes the weighted sum 
\begin{equation}
    \label{eq:wighted_sum}
    z = \sum_i w_{i} x_{i} + b \, ,
\end{equation}
where the set $\{ w_{i}\}$ contains the weights at the given layer, while $b$ is an offset known as bias. Subsequently, a nonlinear function $f$, known as \emph{activation function}, is applied to the result $z$, yielding $y=f(z)$. In this work, we will deploy the standard sigmoid function, defined as $f(z) = 1/(1+e^{-z})$. The outputs of each layer serves as the input of the next layer, thus, layer after layer, the output data become an increasingly more complex function of the input data. After propagating through a finite number of layers, we get to the final \emph{output layer}, whose specific structure responds to the task we need to perform, as it will become clearer later. A sketch of the setup is provided in Fig.~\ref{fig:sketch}. The whole point of training a NN is to optimise the array of free parameters made of weights and biases. 
Moreover, in order to define the problem we aim to tackle using NNs, we need to specify the corresponding \emph{loss function}, as we will discuss in the following sections.

\subsection{Classification}
\label{sec:classificatin}

The first task we want to tackle is a ternary classification problem. In other words, we would like to use ML to classify trajectories corresponding to one of the classes of SDs previously introduced, i.e. Ohmic, sub-Ohmic, and super-Ohmic. To this end, the output layer contains as many artificial neurons as the total number of classes $N_c$, i.e., $N_c=3$, since each neuron should be responsible for a class. In fact, it is supposed to represent the likelihood that each trajectory falls into a given class.  By calling $z_j$ the value calculated by the $j$-th neuron in the output layer through the linear combination of the previous layer's values, the corresponding output value $y_j$ is obtained as~\cite{Marquardt:2021}
\begin{equation}
    \label{eq:softmax}
   \hat{ y}_j = f(z_1, \ldots, z_{N_c}) = \frac{e^{z_j}}{\sum_{k = 1}^{N_c}e^{z_k}} \, ,
\end{equation}
with $j=1,\ldots, N_c$ , where $f$ is the so-called \emph{softmax} activation function. This choice yields a well-defined normalised probability distribution of the output values.

In order to solve our classification problem, we need to suitably define the loss function. A common choice is give by the so-called \emph{categorical cross-entropy}, defined as~\cite{Murphy:2012}
\begin{equation}
    \label{eq:cross-entropy}
    L(\hat{y},y) = - \frac{1}{N_t} \sum_{i=1}^{N_t} \sum_{j=1}^{N_c} y_{ij} \log{\hat{y}_{ij}} \, ,
\end{equation}
where $N_t$ is the total number of trajectories sampled in the dataset, $y_{ij}$ is the true probability that the $i$-th trajectory belongs to the the $j$-th class, while $\hat{y}_{ij}$ is the corresponding predicted probability.

\subsection{Regression}
\label{sec:regression}

As a second task, we perform regression on the relevant parameters defining the SD (\ref{eq:SD_general_s}), i.e. the Ohmicity parameter $s$, the coupling strength $\eta$, and the cut-off frequency $\omega_c$. The network layout is similar to the one used for the classification tasks, except for a few differences. We used the \emph{linear} activation function, defined as $f(z) = z$, which allows the network to produce a wide range of real-valued outputs without restrictions. 
The loss function for this task is the mean squared error (MSE), given by~\cite{Murphy:2012}

\begin{equation}
    \label{eq:MSE}
    L(\hat{y},y) = \frac{1}{N_t} \sum_{i=1}^{N_t}  (y_i - \hat{y}_i)^2 \, ,
\end{equation}
where $y_i$ and $\hat{y}_i$ are the actual and the predicted values for the $i$-th trajectory, respectively.  
The number of artificial neurons in the output layer is fixed by the number of parameters we want to determine, i.e. up to three, as the parameters entering in Eq.~(\ref{eq:SD_general_s}) are $s, \eta, \omega_c$.

\section{Analysis and results}
\label{sec:results}

In this Section, we present the results of our numerical experiments both for the classification and regression tasks. First, we need to solve the dynamics of the system in a given interval of time $[t_{\rm min},t_{\rm max}]$, which allows us to track the expectation value $\langle O(t) \rangle$. In particular, we focus on $\langle \sigma_x(t) \rangle$. This choice does not affect the generality of our argument; other choices of the observable are indeed possible, as long as the corresponding expectation value displays a non-trivial dependence on the SD. For instance, this is not the case for $\langle \sigma_z \rangle$ for the pure dephasing dynamics described in Sec.~\ref{sec:setting} . We generate a dataset containing the expectation values $\langle \sigma_x(t) \rangle$ corresponding to different choices of the SD [cf. Eq.~(\ref{eq:SD_general_s})] thus ensuring that the three classes
-- sub-Ohmic, Ohmic, super-Ohmic -- are equally represented. 
We numerically sample $\langle \sigma_x(t) \rangle$ at $N$ times $t_n$, with $n=0,\ldots, N-1$ in the given interval $[t_{\rm min},t_{\rm max}]$. 
We generate a \emph{training set} containing $N_{\rm train}$ curves to train the model, a \emph{validation set} consisting of $N_{\rm valid}$ curves that are used to monitor the performance during training, a \emph{test set} with $N_{\rm test}$ curves used to assess the accuracy of the fully trained network.
Before feeding the NN with our data, we need to Fourier transform the time series $\{\langle \sigma_x (t_n) \rangle\}_{n=0,\ldots, N-1}$. By calculating the discrete Fourier transform through Eq.~(\ref{eq:Fourier_coeff}), we obtain the Fourier coefficients $X_k$ with $k=0, \dots, N-1$, which are in general complex numbers. Therefore, each curve sampled at $N$ points yields $N$ Fourier coefficients, which result in their $2N$ real and imaginary parts. 
We feed the real and imaginary parts of the Fourier coefficients $X_k$ to an artificial NN, therefore the number of neurons in the input layer is $2 N$. The specific architecture of the network's inner layers is iteratively refined, adding layers and neurons until the network achieves high accuracy without overfitting. The number of neurons in the output layer is fixed by the task. For classification, we have $3$ artificial neurons, each corresponding to one of the classes, i.e. sub-Ohmic, Ohmic, super-Ohmic. For regression, the number of artificial neurons coincides with the number of parameters in Eq.~(\ref{eq:SD_general_s}), which fix the form of the SD, i.e. $s, \eta, \omega_c$.

In our numerical simulations, we consider $t_{\rm min} = 0$ and $ t_{\rm max} = 10$ in units of $\omega_0$, and  sample the trajectories at $N=400$ points. Unless otherwise stated, the NN for each model consists of the input layer followed by $2$ hidden layers where the first (second) hidden layer comprises 250 (80) neurons.
We optimise the NNs using whole batch gradient descent and the Adam optimiser with a learning rate of $1 \times 10^{-4}$. Without loss of generality, we start from the initial state $\rho(0)=| + \rangle \! \langle + |$, with $\ket{+} = (\ket{0} + \ket{1})/\sqrt{2}$, and  focus on the case of a zero-temperature bath. This does not limit the conclusions of our studies. In particular, the choice of a non-zero temperature would only affect the dynamics of the system, as the oscillations would persist for longer intervals of time --- cf. Fig.~\ref{fig:deph_curves}.  We generate a training, validation, and test-set of size $N_{\rm train}/2 = N_{\rm valid} = N_{\rm test} = 2400$. The code employed for data generation, the datasets, and the code utilized for subsequent analysis are available in  GitHub respositories~\cite{github1,github2}.

\subsection{Classification}
We first tackle the classification task, exploring two different scenarios. The former, where we let the Ohmicity parameter $s$ vary, while we keep the cut-off frequency $\omega_c$ and coupling strength $\eta$ fixed. The latter where we let the three parameters $s,\omega_c,\eta$ vary in their respective intervals.

In the former case, we take $\eta = 0.25$, $\omega_c = 0.5$. We try to train the network to tackle increasingly challenging tasks. At first, we consider the case where the three Ohmicity classes are well separated. To this end, we construct our dataset by choosing $s$ within the intervals $ \left( 0, 0.5 \right]$ and $\left[ 1.5, 4 \right]$ for sub-Ohmic and super-Ohmic, respectively. If the SD is Ohmic then $s = 1$. Since there is a neat separation between the different classes, we can anticipate that the performance of the NN will be relatively high. Our numerical experiments confirm this expectation, as the accuracy of the network evaluated on both the training and the test set reaches $100\%$ after approximately $ 80$ iterations of training. 

The classification task becomes a bit more difficult when we choose $s$ in the intervals $(0, 1)$ and $(1,4]$ for the cases of sub-Ohmic or super-Ohmic dissipation, respectively. This choice has a direct consequence on the dynamics, thus on $\langle \sigma_x (t)\rangle$, which would have a similar behaviour over time, even when they correspond to different Ohmicity classes. Despite the fact that the differences between the curves corresponding to different classes are less pronounced, we still obtain a good accuracy in our numerical experiments. The final network training accuracy in this case reaches $99.31 \% $ after around $5000$ training iterations, while the final test accuracy reaches $99.50 \%$.

\begin{figure}[h]
\centering
\includegraphics[width=0.8\textwidth]{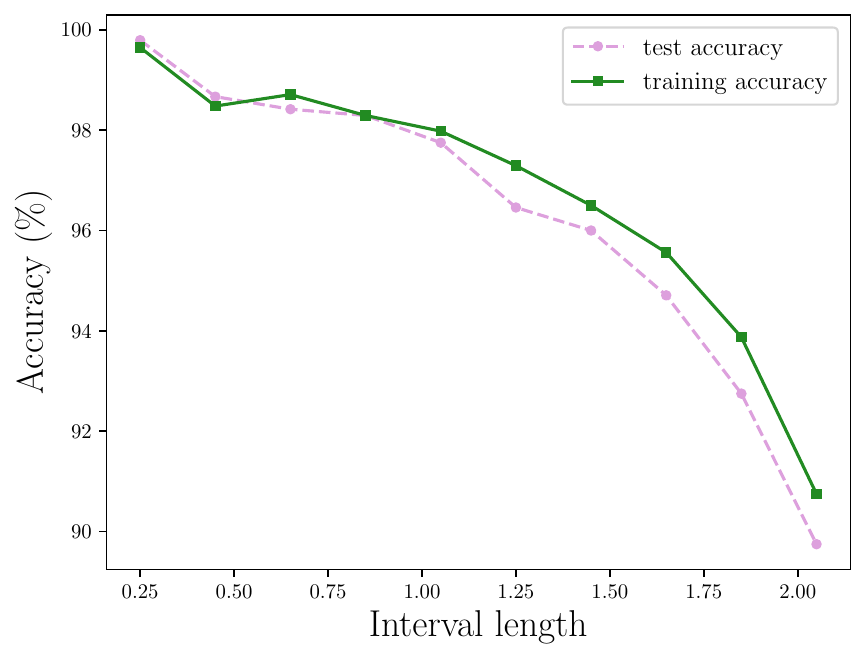}
\caption{Classification task for pure dephasing dynamics. We plot the classification accuracy against the length of the interval from which $\eta$ and $\omega_c$ are sampled.}
\label{fig:dephasing_accuracy}
\end{figure}

We can then make the task even more challenging, allowing not only $s$ to vary, but also $\eta$ and $\omega_c$, so that we can assess the performance of the training as we increase the length of the interval from which they are sampled. We assume that the Ohmicity parameter $s$ is taken within the intervals $(0, 1)$ and $(1,4]$ for sub-Ohmic or super-Ohmic SDs, respectively. Initially, we set both $\eta$ and $\omega_c$ equal to a given value of $0.25$, then we let them vary into the interval $\left[ 0.25, 0.45 \right]$. We increase the upper bound in increments of $0.2$ until the interval becomes $\left[ 0.25, 2.05 \right]$. In this scenario, the differences between the three classes are more subtle, which make the classification task significantly more difficult. The classification results after $2 \times 10^{4}$ training iterations are shown in Fig.~\ref{fig:dephasing_accuracy}, where the solid line is associated with the accuracy evaluated on the training set, while the dashed curve is associated with the accuracy evaluated on the test set. As expected, we can see that the accuracy decreases as we consider larger intervals for $\eta$ and $\omega_c$, due to a larger amount of noise in the dataset.

Some comments are now in order. First, we should mention that qualitatively similar results are obtained for the same class of SDs [Cf.~Eq.(\ref{eq:SD_general_s})] when one considers quantum systems governed by different dynamics. In Ref.~\cite{barr2024spectral}, a different model inducing a non-Markovian amplitude damping mechanism is analysed. The latter emerges when the quantum system is coupled to the environment through $X = \sigma_x/2$ in Eq.~(\ref{eq:H_int}). The specific form of the SD allows for an analytic derivation of a master equation, accurate up to the second order in the coupling constant~\cite{Breuer:2001,Clos:2012}. Although the dynamics is physically richer, one can still achieve very good accuracy in classifying the curves according to the corresponding SD.
Second, it is worth emphasising that the accuracy achieved in the classification task is inherently impacted by various sources of noise, e.g. sampling noise that inevitably arises in experimental scenarios. In Ref.~\cite{barr2024spectral} the influence of sampling noise is discussed, showing that increasing noise levels have an impact on the performance of the NN, as they hamper generalization to previously unseen data.

\subsection{Regression}
\label{sec:results_deph_regression}

For regression experiments, we again consider two scenarios: one where $\omega_c$ and $\eta$ are fixed, with only $s$ varying, and the other where all three parameters are allowed to vary. 
In the first scenario, we only estimate the value of $s$, as it is the only parameter that varies. We initially train a model using the dataset where the values of $s$ are clearly separated across the sub-Ohmic, Ohmic and super-Ohmic classes. Specifically, the dataset used here is constructed in the same way as the one used for the classification task. %the sub-Ohmic and  super-Ohmic  SDs are taken in the intervals $(0, 0.5]$ and $[1.5, 4]$, while the Ohmic case is defined by $s=1$. The coupling strength is fixed at $\eta = 0.25$, and the cut-off frequency is set to $\omega_c = 0.5$. 
Given the distinct changes in the trajectories as $s$ varies, we expect that regression of $s$ will be relatively easy for the model. 

\begin{figure}
\centering
\subfloat[]{\label{fig:predictedsvreal_separated}{\includegraphics[width=0.48\textwidth]{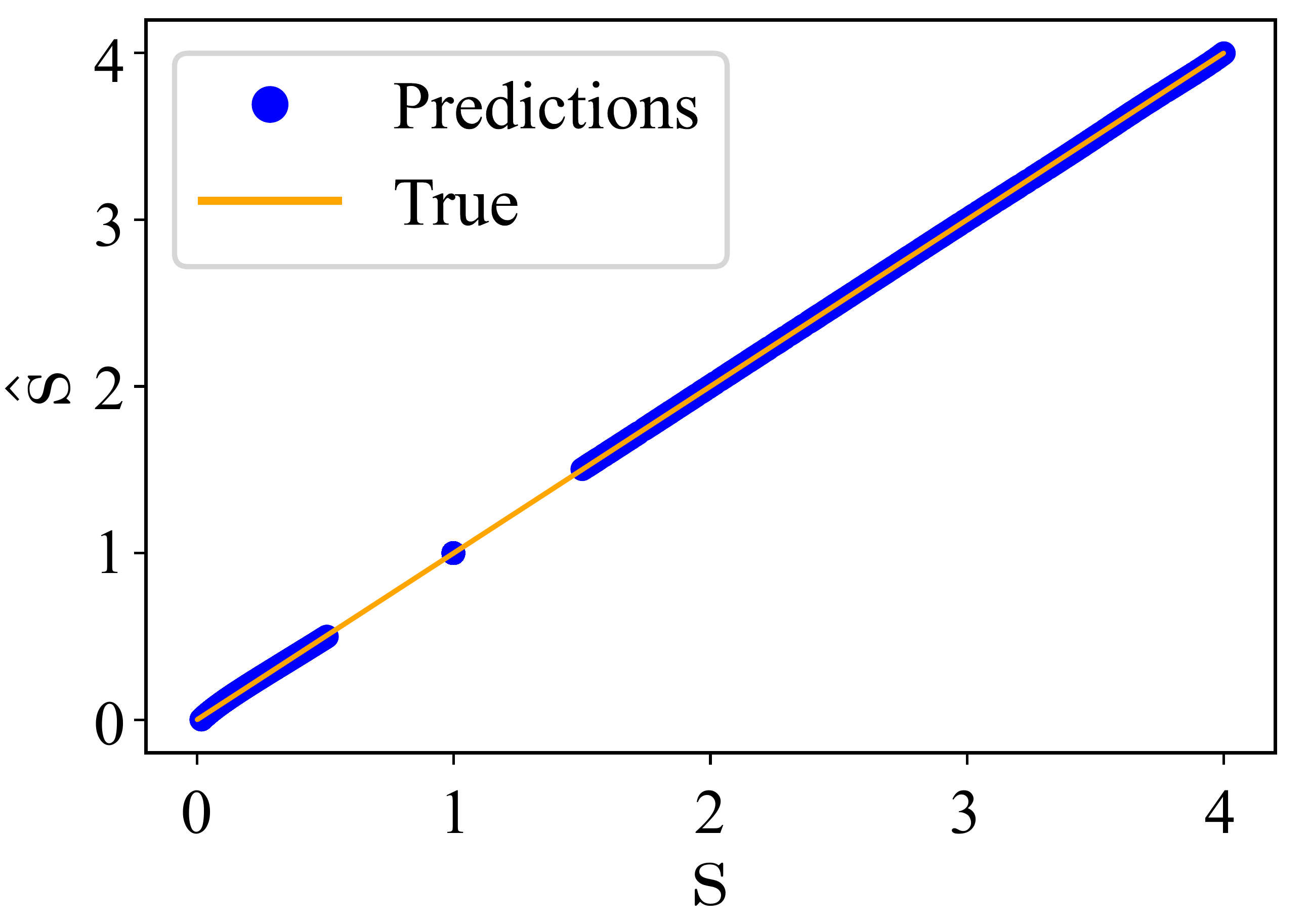}}}
\subfloat[]{\label{fig:predictedsvreal_notseparated}{\includegraphics[width=0.48\textwidth]{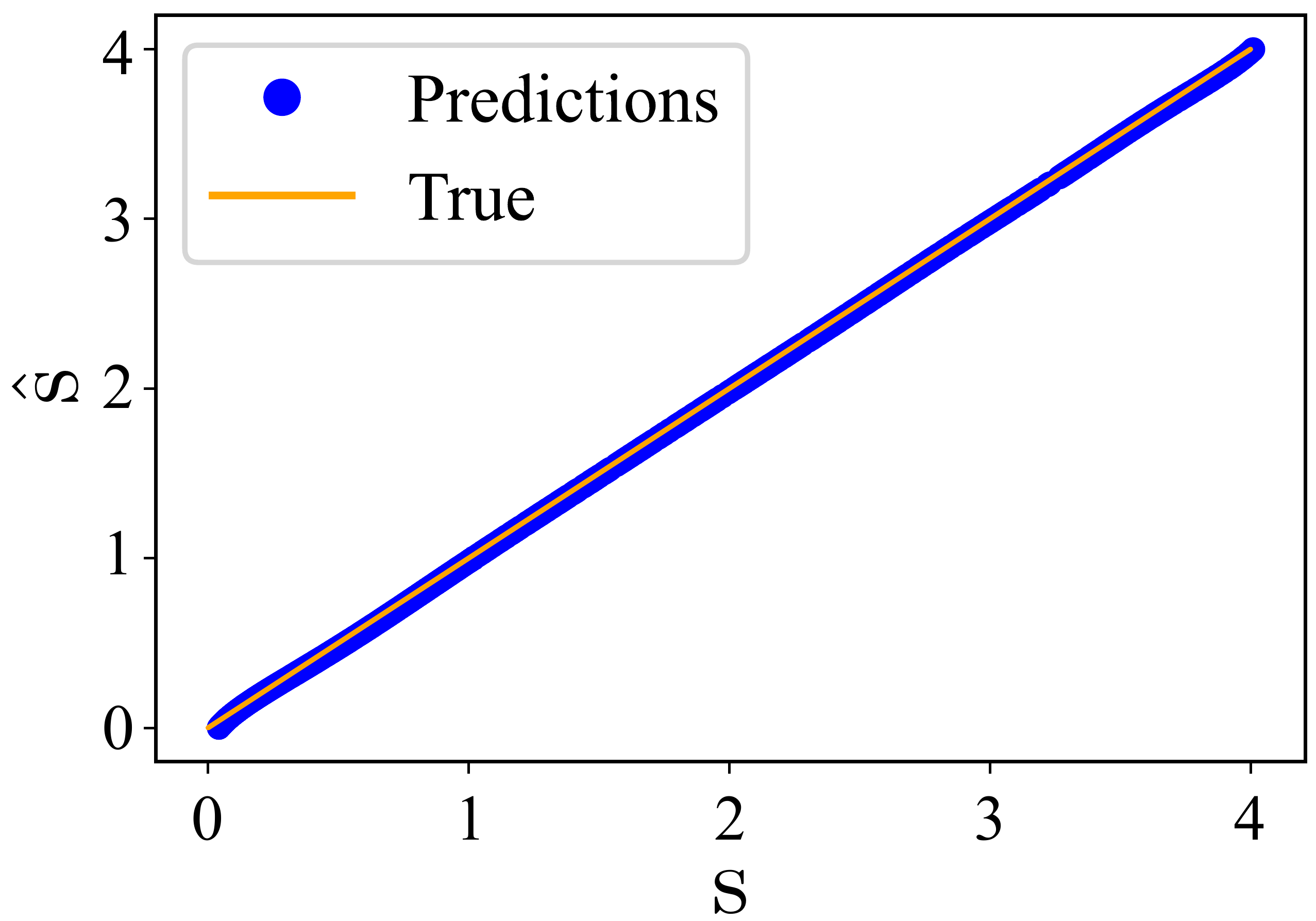}}}
\hfill
\subfloat[]{\label{fig:errorvs_separated}{\includegraphics[width=0.48\textwidth]{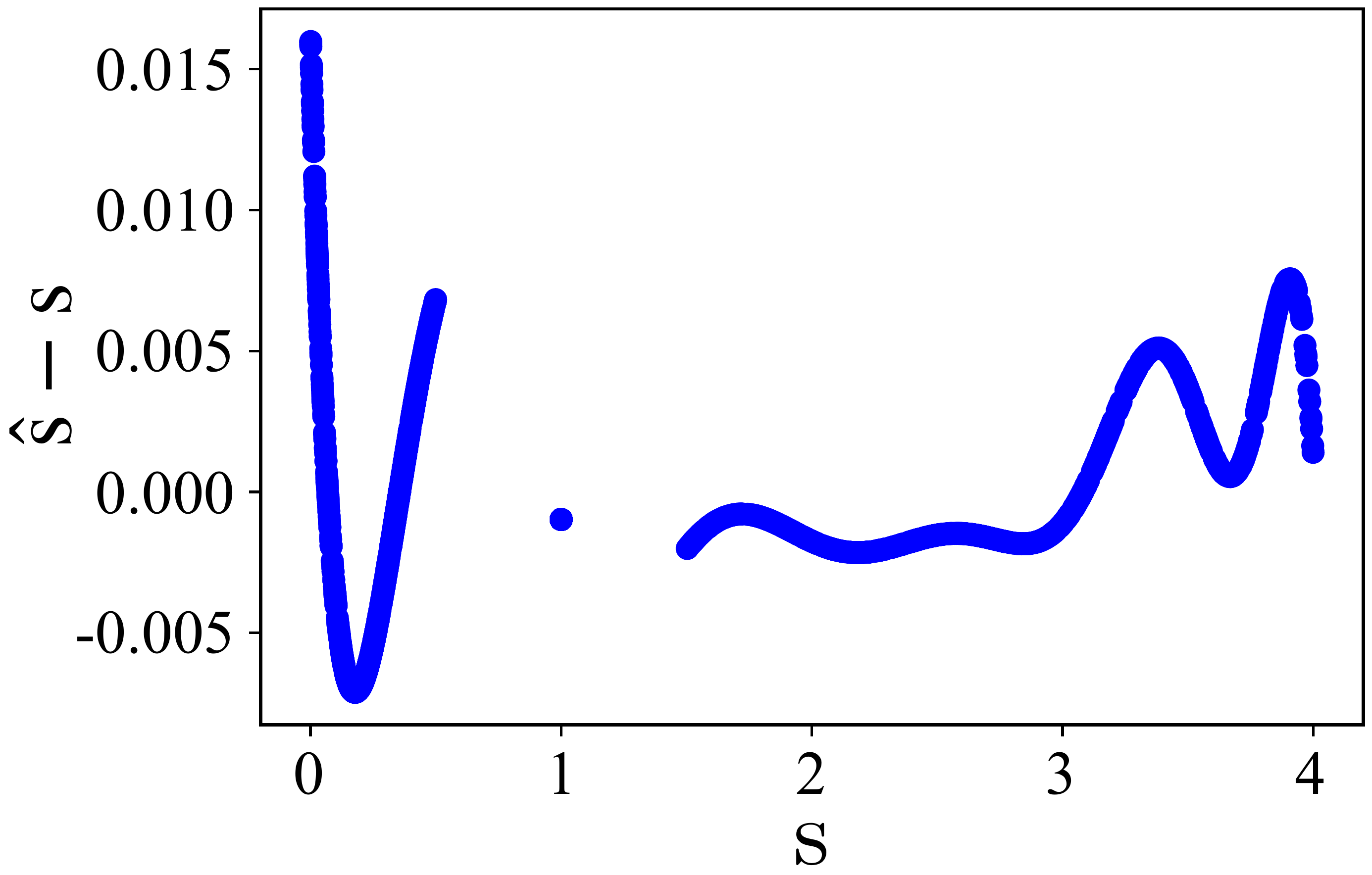}}}
\subfloat[]{\label{fig:errorvs_notseparated}{\includegraphics[width=0.48\textwidth]{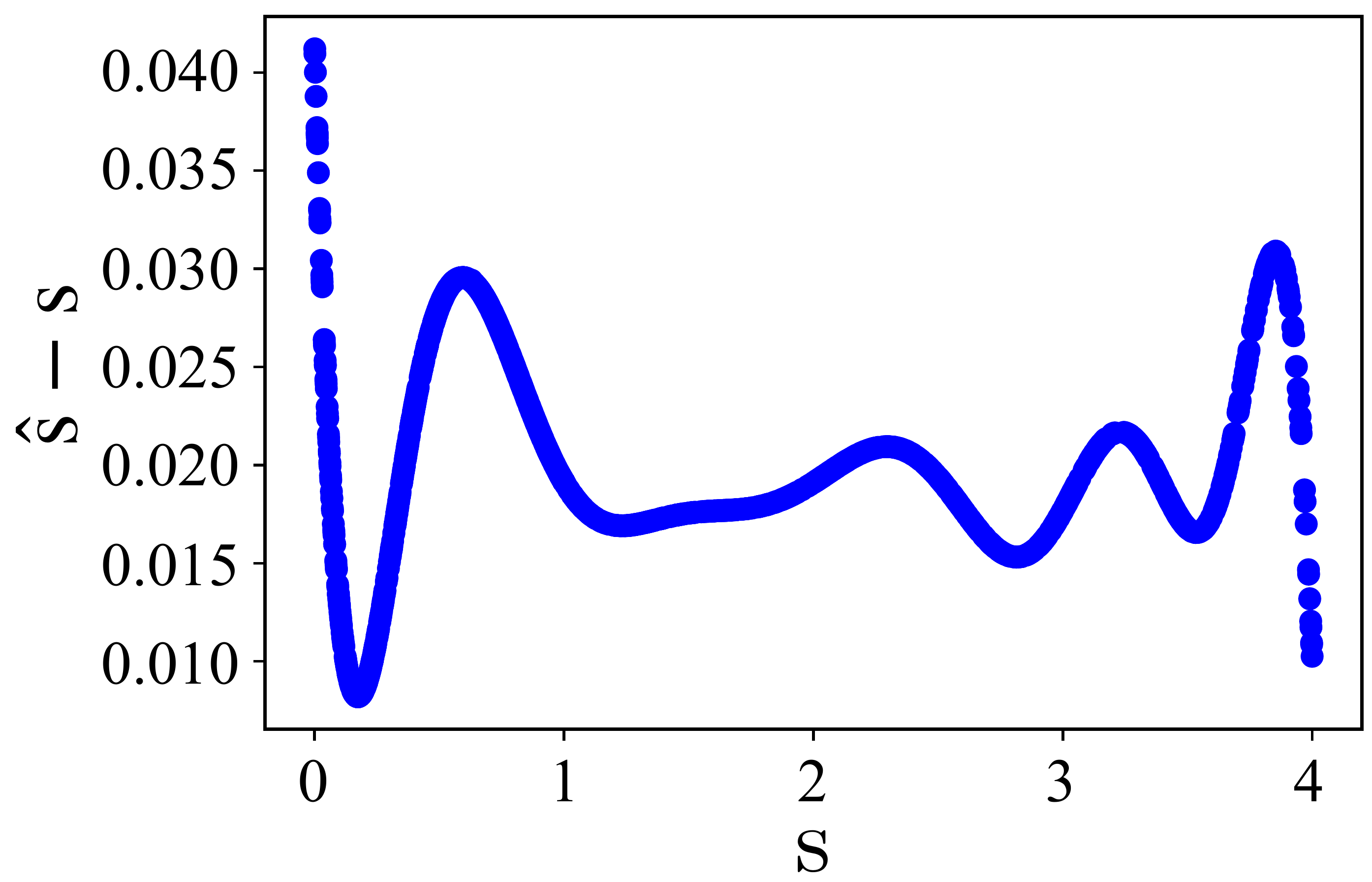}}}
\hfill
\subfloat[]{\label{fig:barchart_separated}{\includegraphics[width=0.5\textwidth]{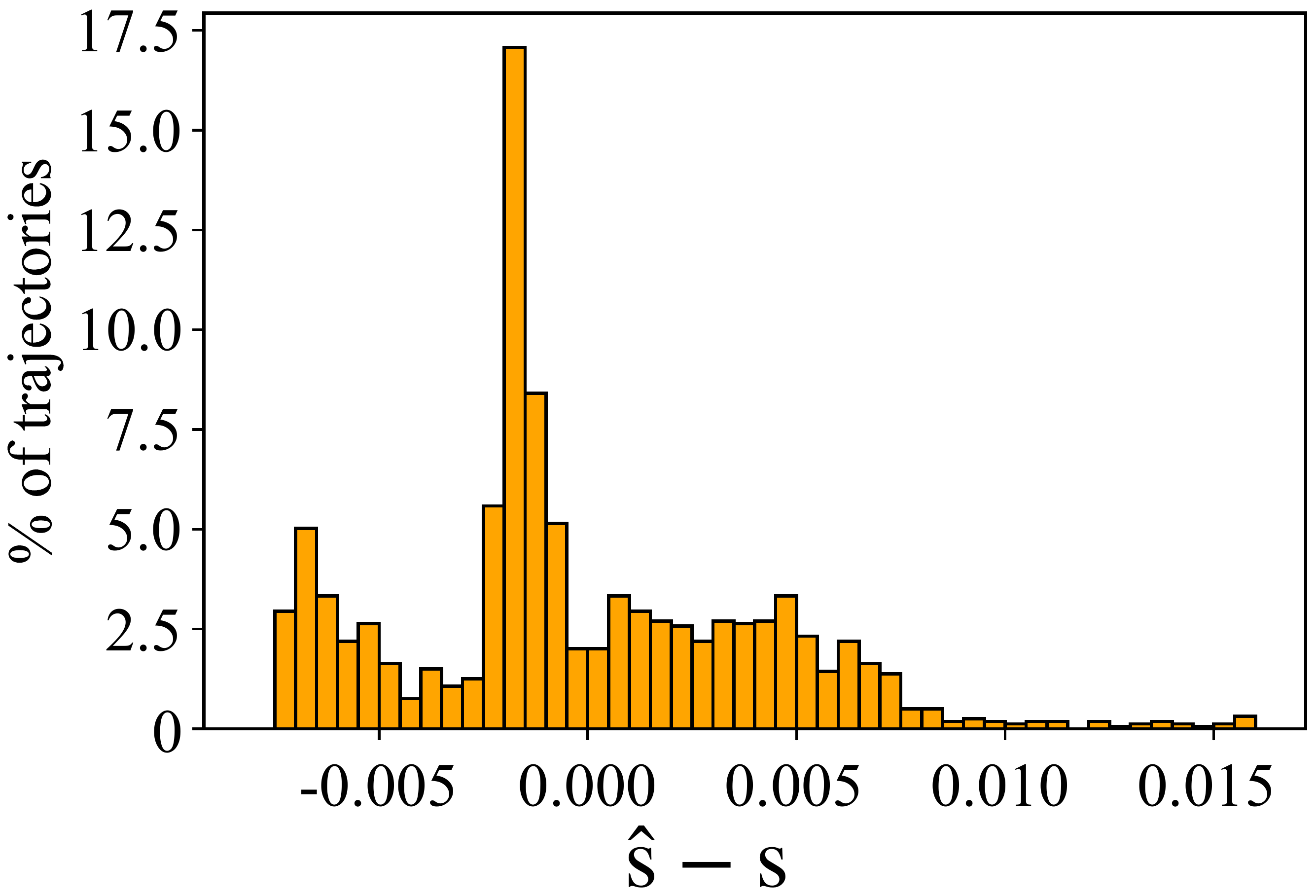}}}
\subfloat[]{\label{fig:barchart_notseparated}{\includegraphics[width=0.48\textwidth]{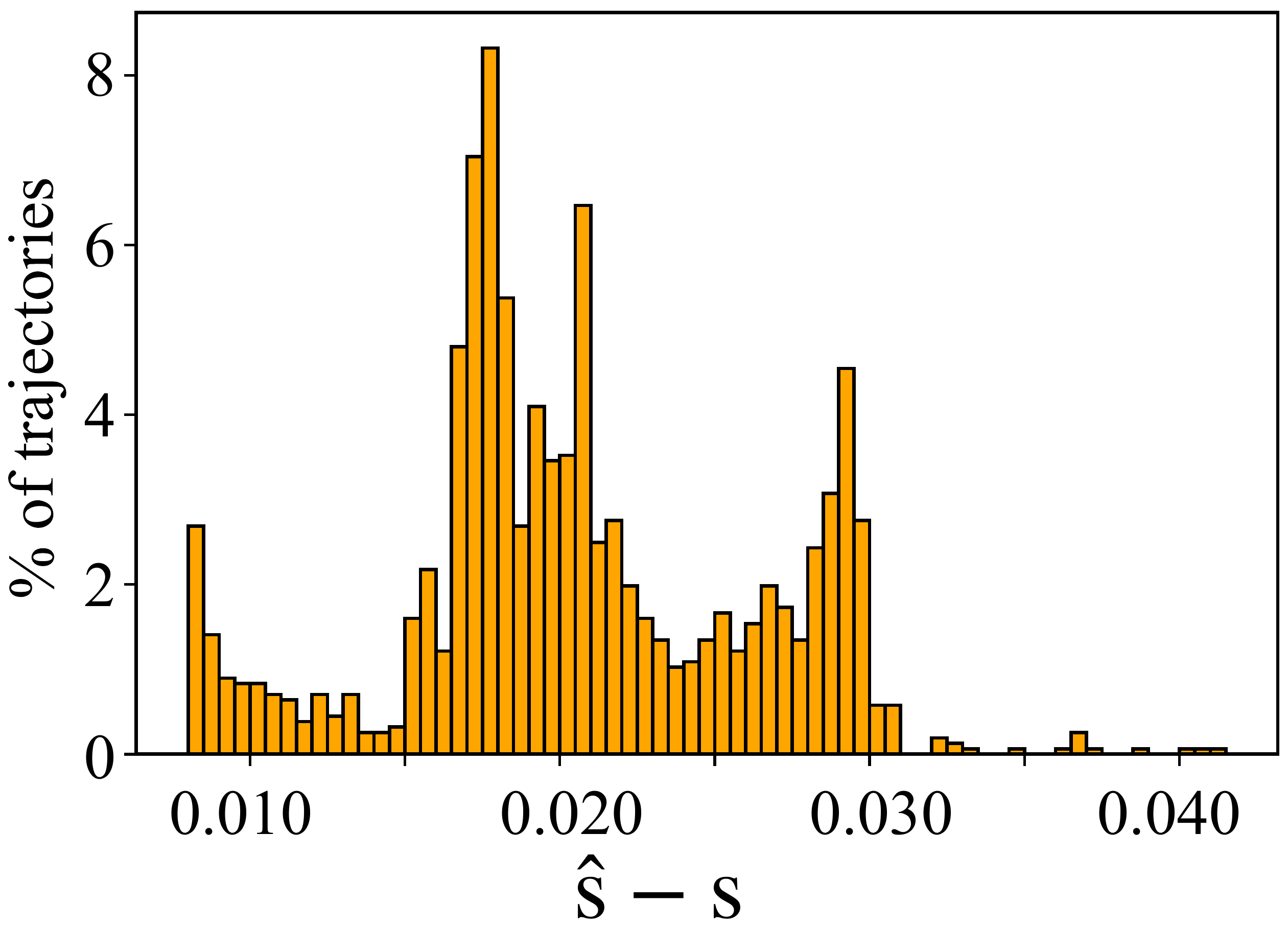}}}
\caption{Pure dephasing model: regression results for the dataset where $\eta = 0.25$ and $\omega_c = 0.5$ are fixed, while the Ohmicity parameter $s$ varies. Panels (a)-(c)-(e) refers to the case where $s \in (0, 0.5)$ if the SD is sub-Ohmic and $s \in [1.5, 4]$ if the SD is super-Ohmic, while Panels (b)-(d)-(f) refer to the case where $s$ belongs to the intervals $(0, 1)$ or $(1, 4]$ for sub-Ohmic or super-Ohmic SDs, respectively. Panels (a)-(b) show the predicted values of $s$ plotted against the true values for the test set, with a reference line indicating perfect predictions. Panels (c)-(d) present the errors (the difference between the predicted and true values) plotted against the true value of $s$. Panels (e)-(f) display a bar chart indicating the percentage of test trajectories for which the errors fall within specific intervals. Only one Ohmic trajectory from the test set is included in the bar chart to prevent skewing the distribution.}
\label{fig:regressionresults_s}
\end{figure} 

After $10^3$ training iterations, we evaluate the performance of the NN on both the training and test sets. The final MSE on the training set is $6.492 \times 10^{-4}$, while the MSE on the test set is $6.507 \times 10^{-4}$. The low and nearly identical MSE values for both sets indicate that the model generalised well and did not overfit. In Fig.~\ref{fig:predictedsvreal_separated} we present the predicted values, $\hat{s}$, plotted against the true values for the test set, along with a reference line to guide the eye. The close alignment between the predictions and the true values is evident from the proximity of the points to the reference line. Additionally,~\ref{fig:errorvs_separated} shows the errors (the difference between the true and predicted values) plotted against the true values of $s$. The errors are generally small across the range of $s$, though some slight variation is observed, particularly for lower values of $s$. The bar chart in~\ref{fig:barchart_separated} illustrates the percentage of trajectories in the test set where the error falls within specific intervals. To avoid the distribution from being skewed by the presence of multiple identical Ohmic trajectories, only one Ohmic trajectory from the test set was included in the bar chart. The tallest bars in the bar chart correspond to intervals where the errors more frequently fall, as reflected in the plot of errors against the true values of $s$. These intervals are observed more often in the error plot, accounting for the higher percentage of trajectories falling within these ranges. Overall, the results indicate that the model performed well, with strong generalisation and accurate predictions across the range of $s$ values.

Next we consider the dataset where $\eta = 0.25$, $\omega_c = 0.5$, with $s \in (0,1)$ for sub-Ohmic SDs and $s \in (1, 4]$ for super-Ohmic SDs. This dataset spans a broader range of $s$ values, allowing for a more comprehensive evaluation of the model's ability to predict the Ohmicity parameter across both the sub-Ohmic and super-Ohmic regimes, even though, given the distinct changes in the trajectories as $s$ varies, we expect that the regression task will also be relatively straightforward for this dataset. In this case, the final MSE on the training set after $10^3$ training iterations was $5.010 \times 10^{-3}$, while the MSE on the test set was $5.009 \times 10^{-3}$. The minimal difference between these values indicates that the model generalises well and avoided overfitting. The results obtained for this case are reproduced in Figs.~\ref{fig:predictedsvreal_notseparated}, \ref{fig:errorvs_notseparated}, and \ref{fig:barchart_notseparated}. The overall spread of errors remains relatively narrow, highlighting the model's high performance across the dataset. In summary, the model demonstrated robust generalisation and accurate predictions for this dataset.

We now extend the regression task to the case where $\eta$ and $\omega_c$ also vary, alongside $s$. This adds complexity to the model's task, as it must now predict all $3$ parameters simultaneously. As in the classification task, we assess the performance of the NN as we progressively increase the upper bound of the interval from which $\eta$ and $\omega_c$ are sampled. Initially, $\eta = \omega_c = 0.25$, while $s$ is allowed to vary within $(0, 1)$ for sub-Ohmic SDs and $(1, 4]$ for super-Ohmic SDs. It is worth emphasising that as long as we consider larger and larger intervals in which to sample $\eta$ and $\omega_c$, the differences between trajectories appear more complex and difficult to interpret, making it harder to distinguish the underlying patterns. 

In this case, since the task is more complex than when $\eta$ and $\omega_c$ were fixed, we need to modify the NN architecture, including more hidden layers. Therefore, we use a NN with $6$ hidden layers: the first $5$ hidden layers contain $250$ neurons, while the final hidden layer contains $80$ neurons.

\begin{figure}[h]
\centering
\includegraphics[width=0.8\textwidth]{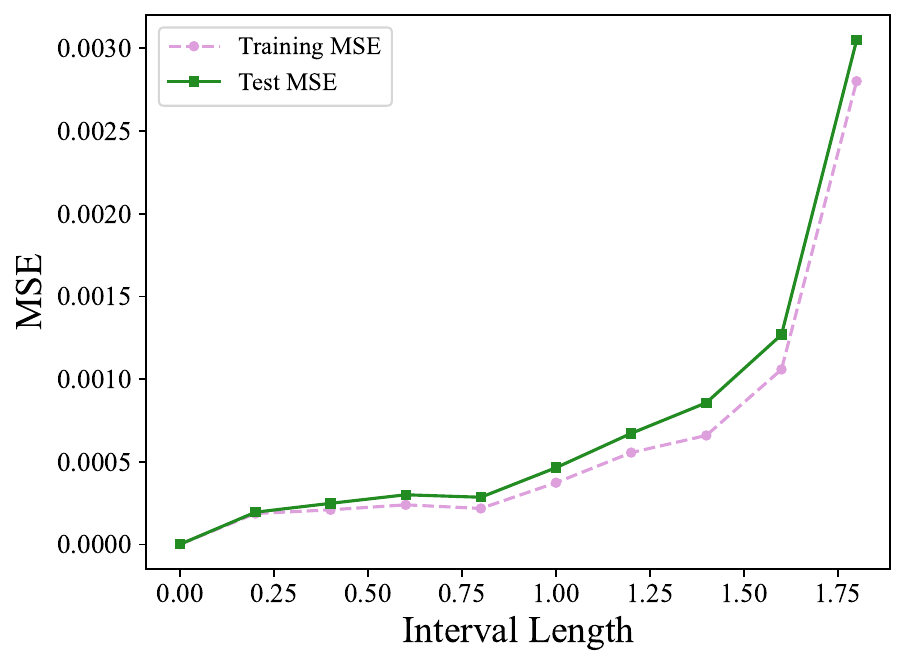}
\caption{Regression task for the pure dephasing model. We plot the MSE against the length of the interval from which $\eta$ and $\omega_c$ are sampled.}
\label{fig:lossvintervallength}
\end{figure}

The results after $1 \times 10^{5}$ training iterations are shown in Fig.~\ref{fig:lossvintervallength}. From Fig.~\ref{fig:lossvintervallength}, we observe that, as the interval length increases, both the training and the test MSE gradually increase. This reflects the increasing difficulty the model faces when predicting across a broader range of parameter values. As $\eta$ and $\omega_c$ are sampled from longer intervals, the complexity of the relationship between the SD parameters and the trajectories increases, making it harder for the model to learn and generalise accurately. As in the classification task, it is worth noting that performance may improve with larger datasets, more training iterations, or exploring a different model architecture.

To gain more insights into the model's performance, we analyse its predictions in more detail for two interval lengths: the shortest non-zero interval length where $\eta, \omega_c \in [0.25, 0.45]$ and the maximum interval length where $\eta, \omega_c \in [0.25, 2.05]$. By comparing these two cases, we aim to better understand how the model's ability to predict $\eta$, $\omega_c$ and $s$ changes as the range of parameter values expands. Specifically, we compare the predicted values of $\eta$, $\omega_c$, and $s$ and with the true values, and analyse the error distribution across the test set.

\begin{figure*}
\centering
\subfloat[]{\label{fig:predictedomegacvreal_0.25and0.45}{\includegraphics[width=0.51\textwidth]{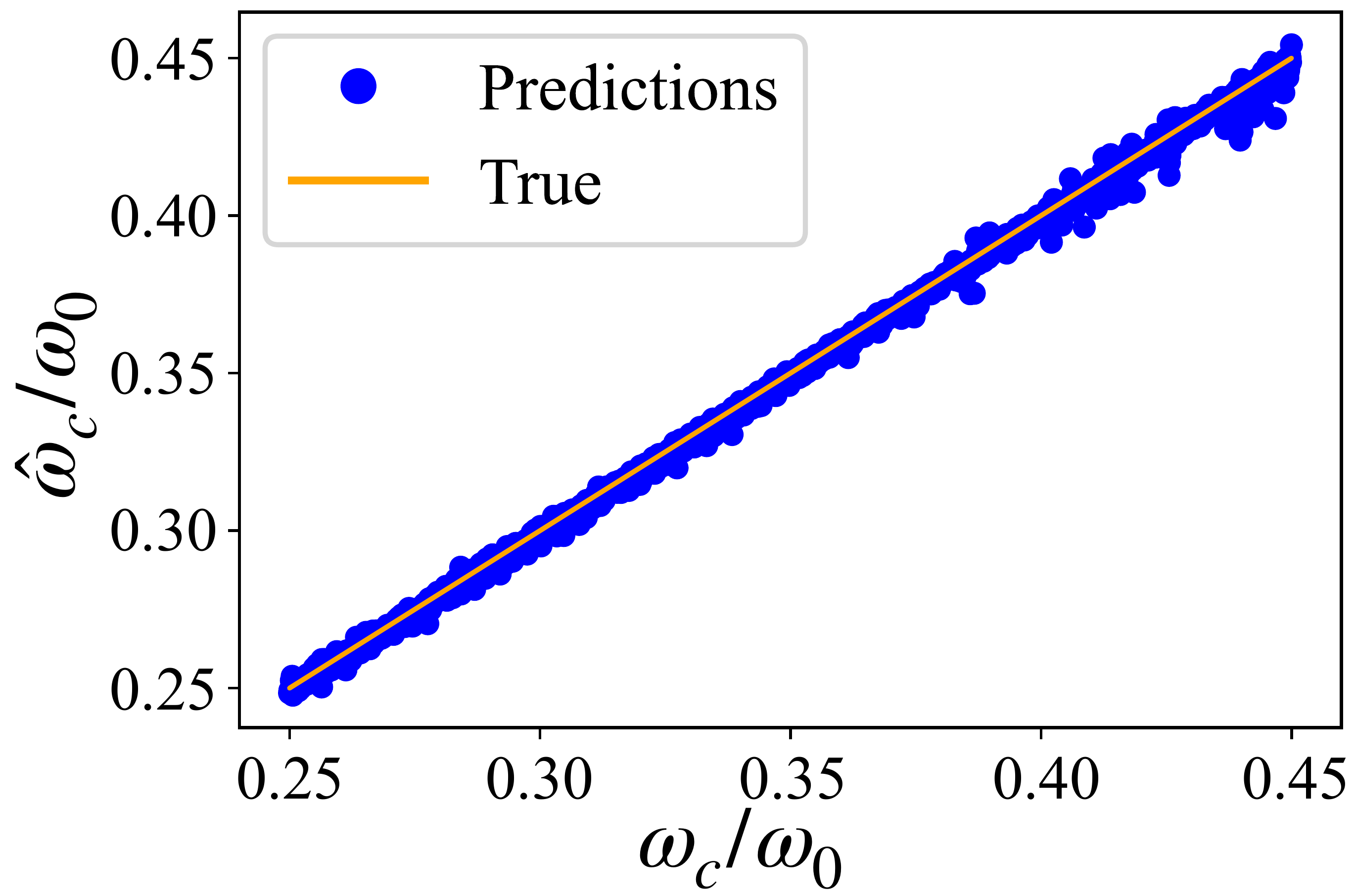}}}
\subfloat[]{\label{fig:omegacbarchart0.45}{\includegraphics[width=0.49\textwidth]{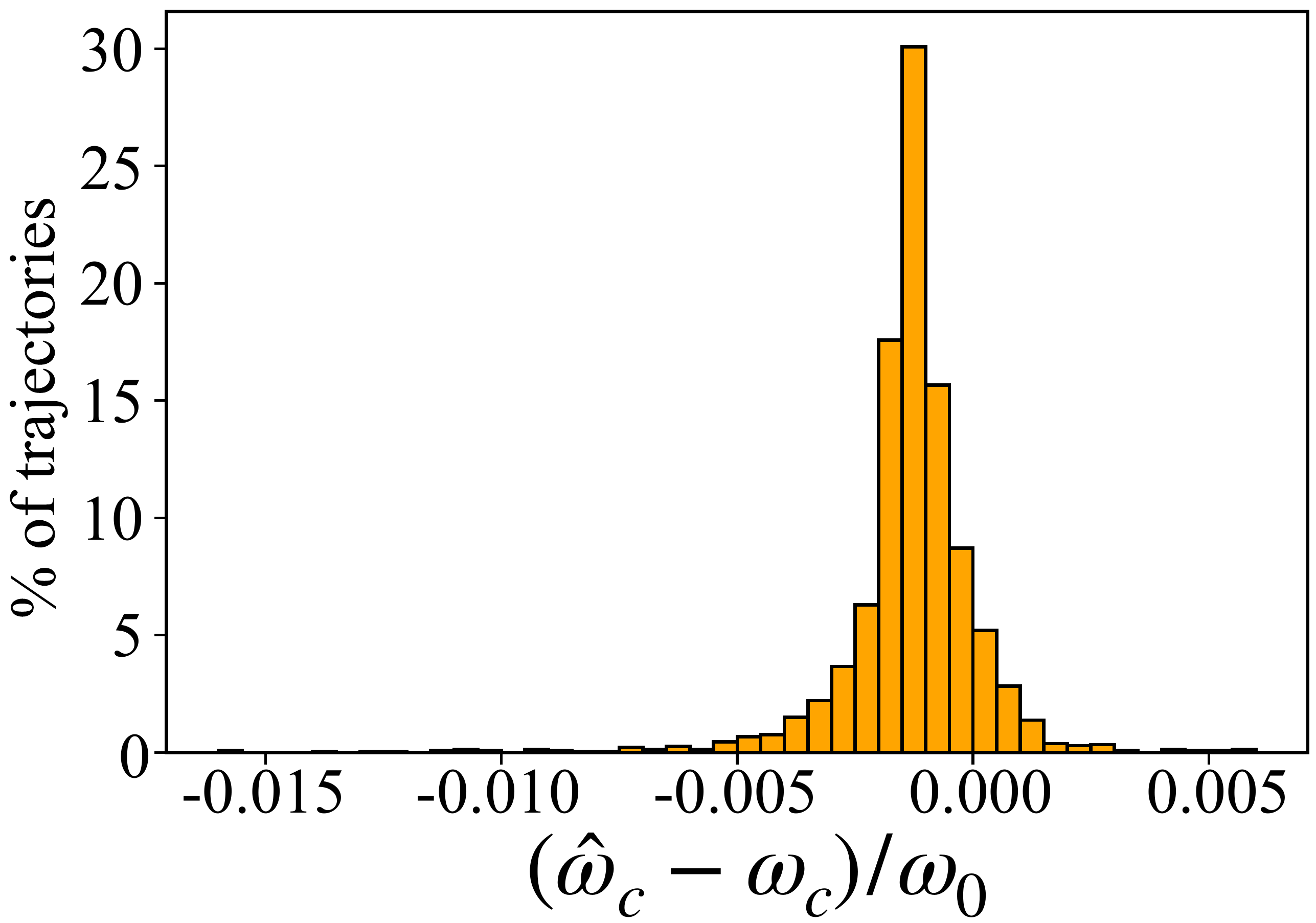}}}
\hfill
\subfloat[]{\label{fig:predictedsvreal_0.25and0.45}{\includegraphics[width=0.49\textwidth]{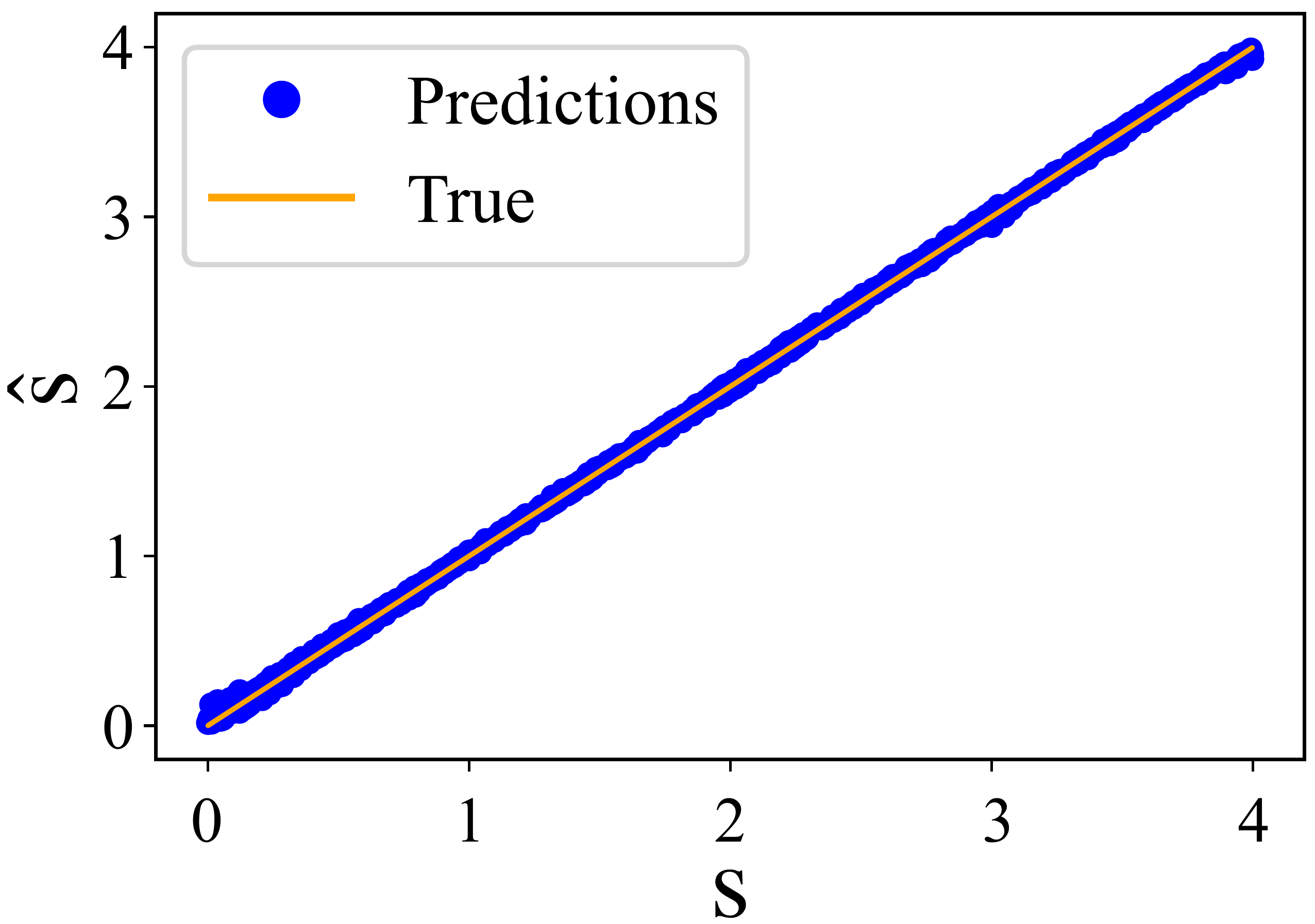}}}
\subfloat[]{\label{fig:sbarchart0.45}{\includegraphics[width=0.49\textwidth]{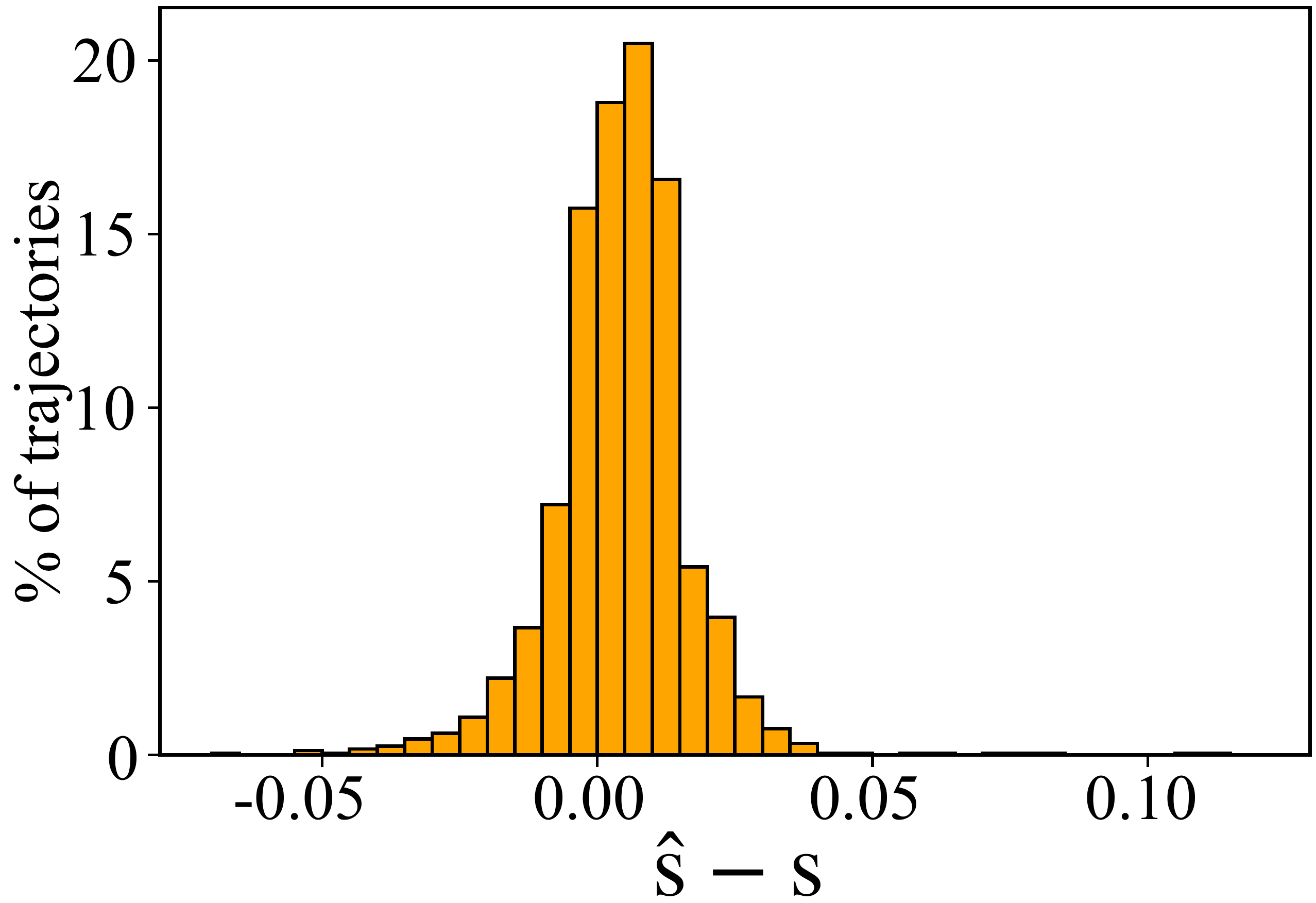}}}
\hfill
\subfloat[]{\label{fig:predictedetavreal0.25and0.45}{\includegraphics[width=0.50\textwidth]{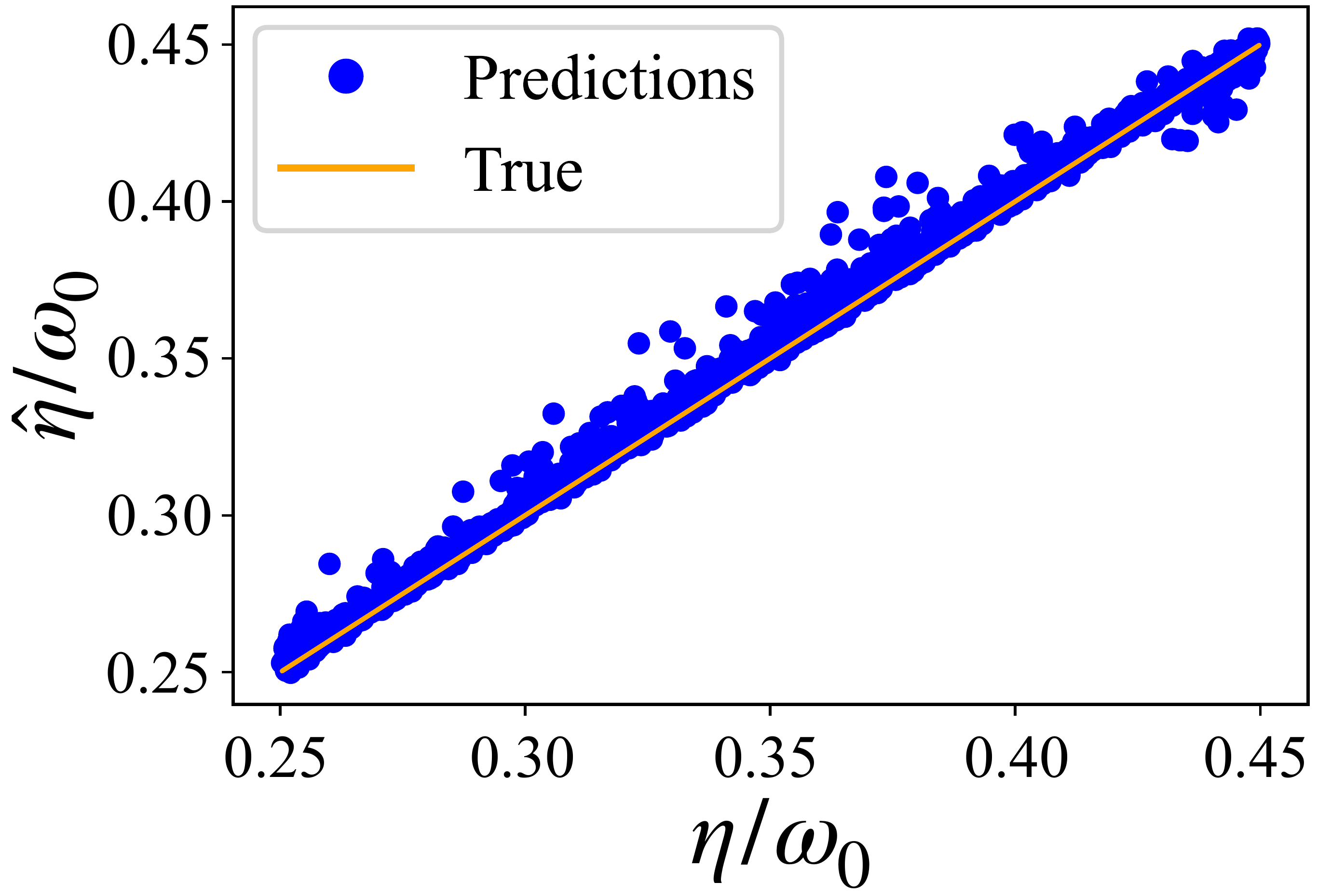}}}
\subfloat[]{\label{fig:etabarchart0.45}{\includegraphics[width=0.49\textwidth]{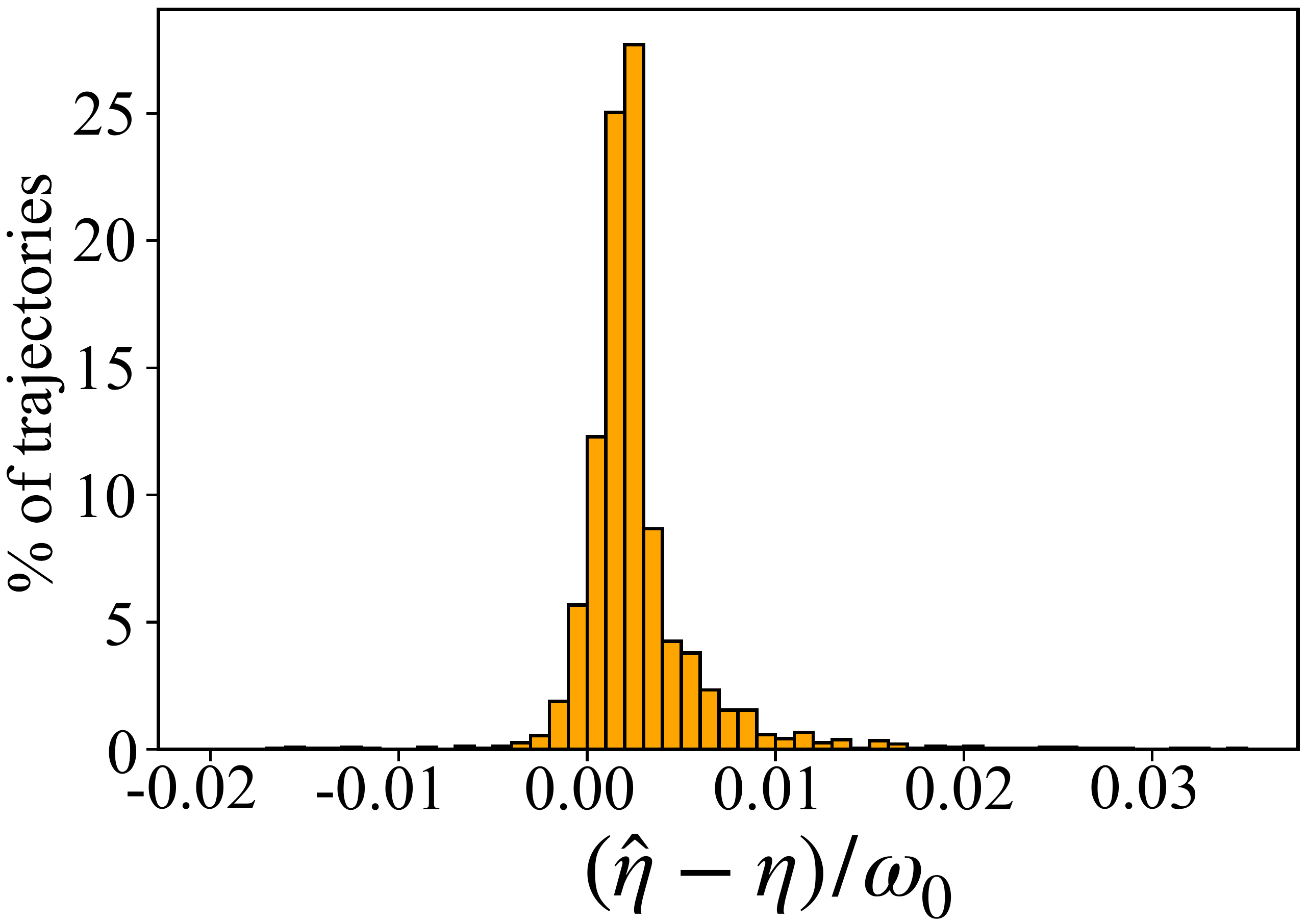}}}
\caption{Pure dephasing model: results of the regression analysis of $\omega_c$, $s$, and $\eta$ for the shortest interval length where $\eta, \omega_c \in [0.25, 0.45]$. Panels (a), (c) and (e) show the predicted values $\hat{\omega}_c$, $\hat{s}$, and $\hat{\eta}$. respectively, against the true values for the test set. Each plot includes a reference line representing the ideal predictions, where the predicted value equals the true value. Panels (b), (d) and (f) show bar charts depicting the distribution of errors between the predicted and true values for $\omega_c$, $s$, and $\eta$, respectively. }
\label{fig:regressionresults_0.25and0.45}
\end{figure*}

For the shortest interval length, where $\eta, \omega_c \in [0.25, 0.45]$, the NN demonstrates strong performance in predicting the parameters $\eta$, $\omega_c$, and $s$. The MSE on the training set is $1.871 \times 10^{-4}$, while the MSE on the test set is $1.952 \times 10^{-4}$. The small difference between these values indicates that the model generalises well and avoids overfitting, suggesting reliable predictions on unseen data. Figs.~\ref{fig:predictedomegacvreal_0.25and0.45}, \ref{fig:predictedsvreal_0.25and0.45}, and~\ref{fig:predictedetavreal0.25and0.45} show the predicted values against the true values for $\omega_c$, $s$ and $\eta$, respectively, for the trajectories in the test set. The corresponding error distributions for each parameter are shown in~\ref{fig:omegacbarchart0.45}, \ref{fig:sbarchart0.45}, and~\ref{fig:etabarchart0.45}.

The plots of the predicted versus true values further illustrate the model's accuracy across all three parameters. For each parameter, the predicted value closely follows the reference line, indicating that the model provides accurate predictions. This consistent accuracy highlights the model's strong performance within this smaller interval length. The error distribution bar charts provide a more nuanced understanding of the model's performance. While the errors for $\omega_c$ and $\eta$ are tightly concentrated around zero, indicating precise predictions, the error distribution for $s$ exhibits a somewhat broader spread. This suggests that although the model performs well overall, it encounters slightly more difficulty in predicting $s$ with the same level of precision as $\eta$ and $\omega_c$. Despite this, the majority of errors for $s$ remain relatively small, reflecting the model's ability to handle the task effectively. In summary, for the shortest interval length, the model is able to predict $\omega_c$, $s$ and $\eta$ with high accuracy and minimal error.

\begin{figure*}
\centering
\subfloat[]{\label{fig:predictedomegacvreal_0.25and2.05}{\includegraphics[width=0.51\textwidth]{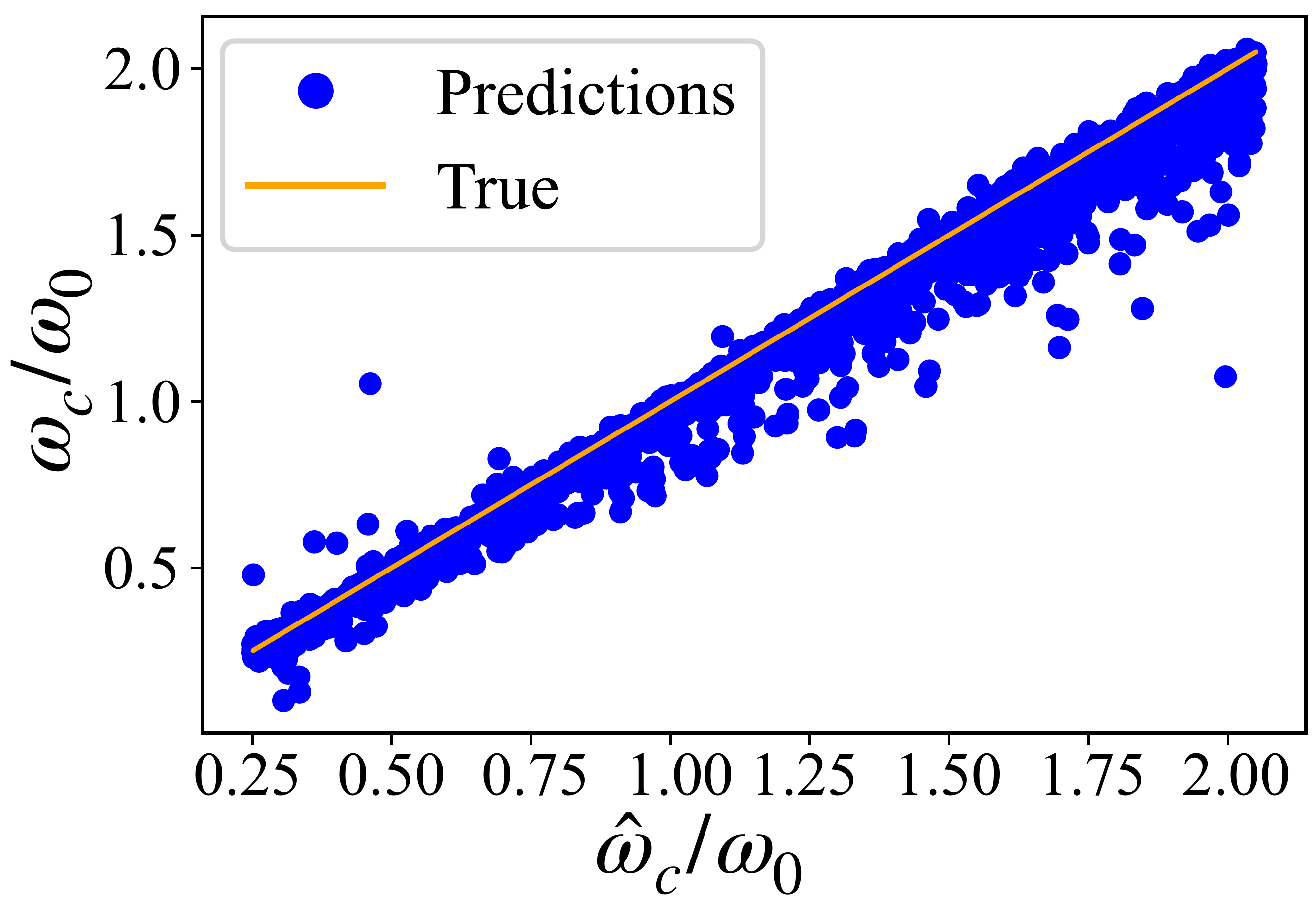}}}
\subfloat[]{\label{fig:omegacbarchart2.05}{\includegraphics[width=0.49\textwidth]{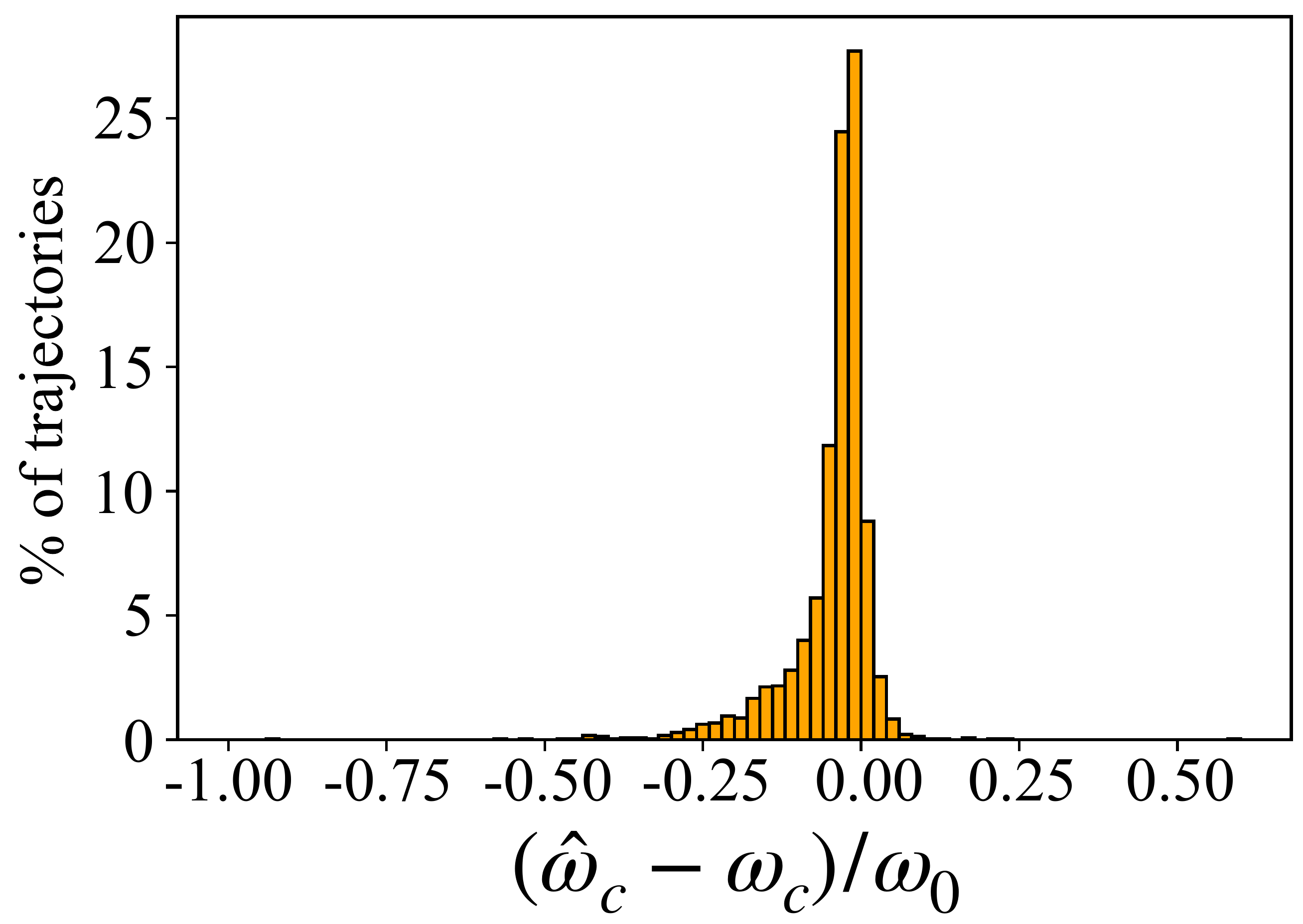}}}
\hfill
\subfloat[]{\label{fig:predictedsvreal_0.25and2.05}{\includegraphics[width=0.49\textwidth]{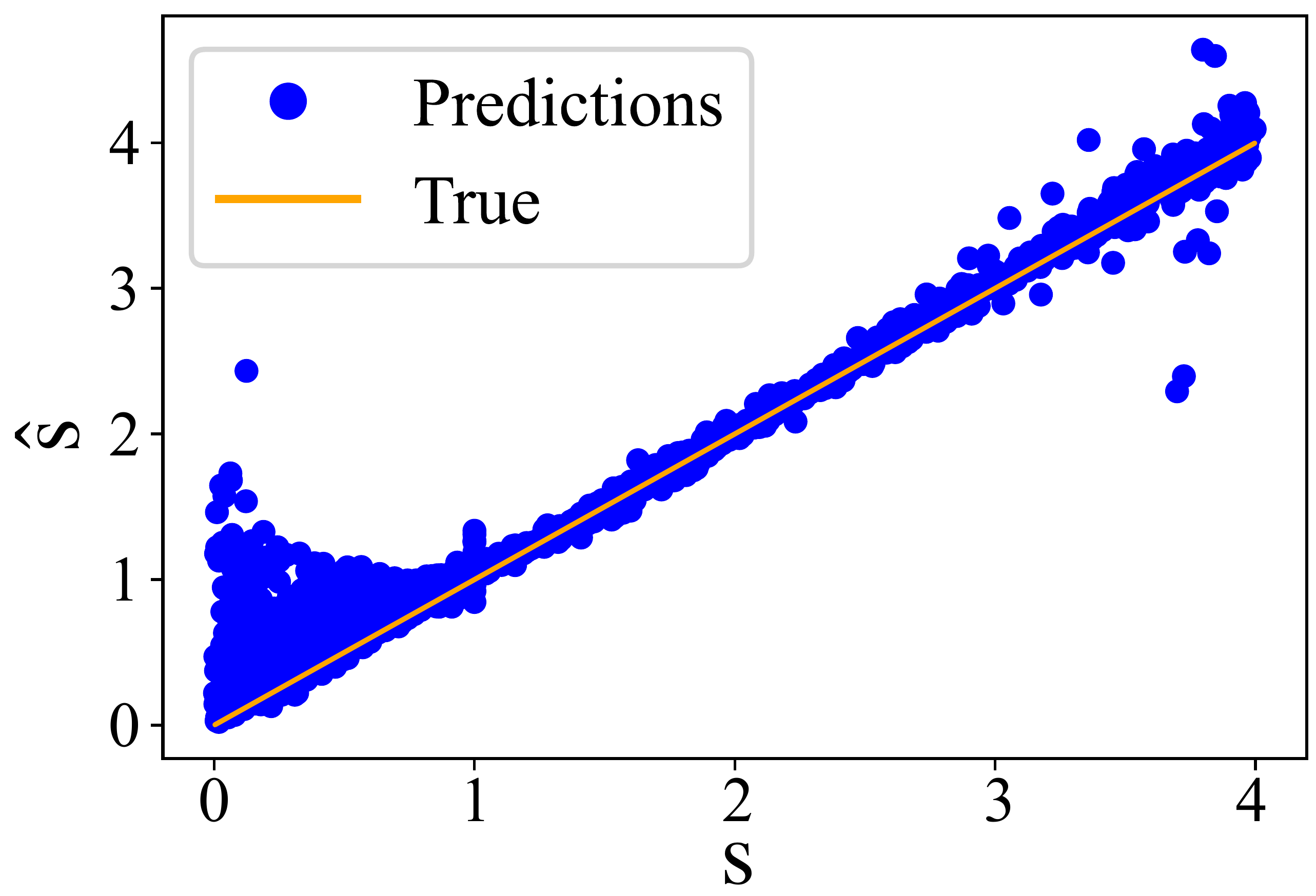}}}
\subfloat[]{\label{fig:sbarchart2.05}{\includegraphics[width=0.49\textwidth]{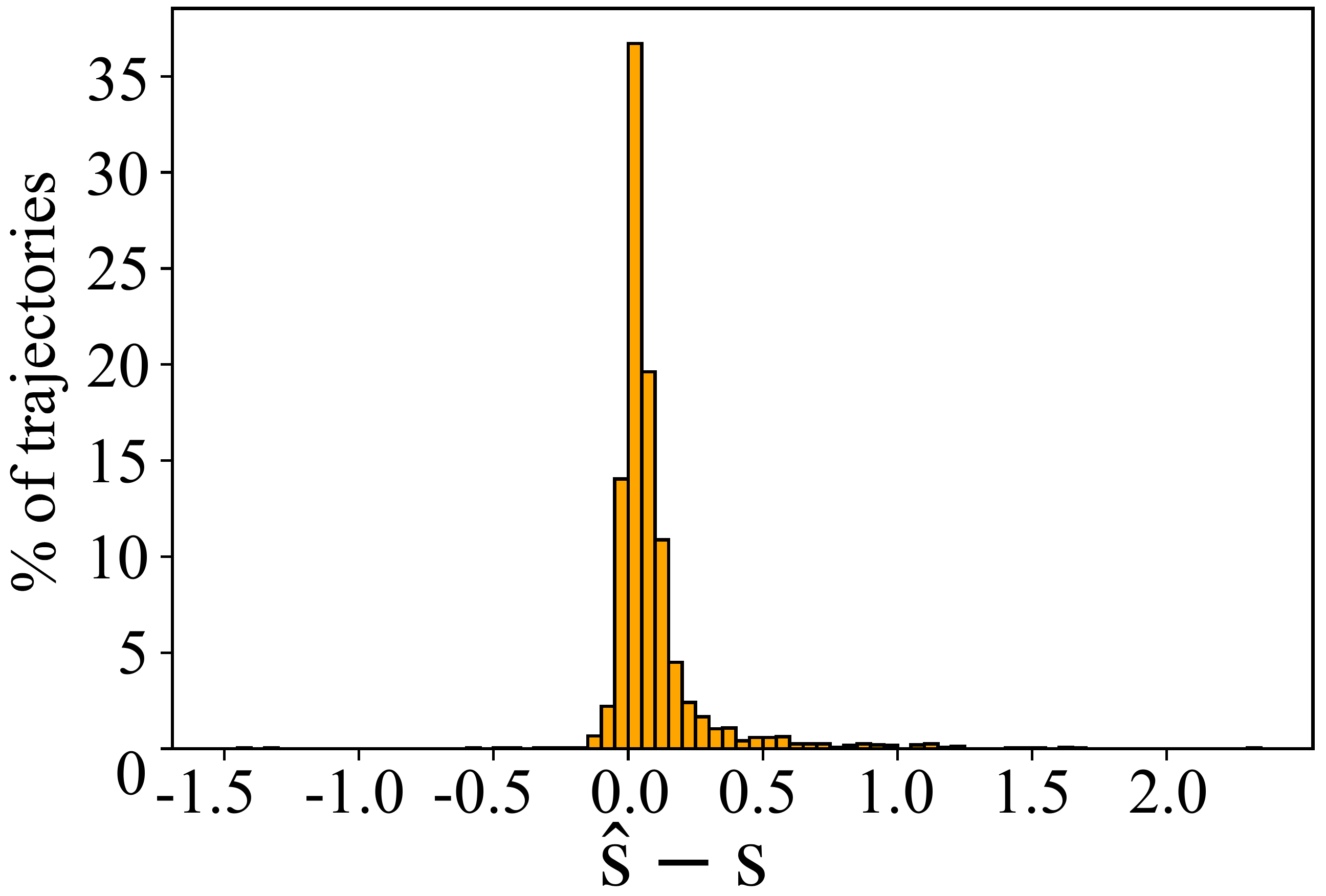}}}
\hfill
\subfloat[]{\label{fig:predictedetavreal0.25and2.05}{\includegraphics[width=0.50\textwidth]{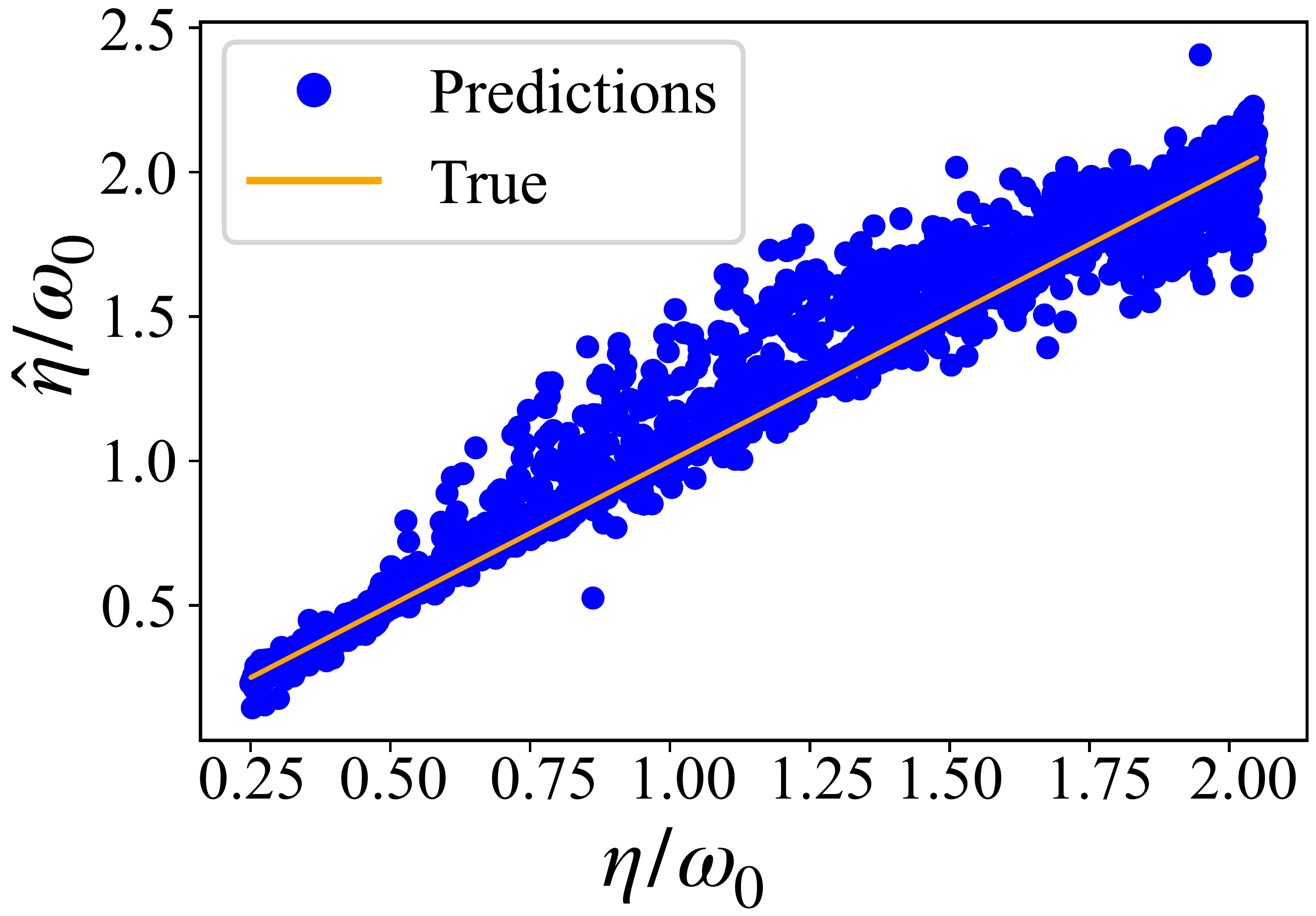}}}
\subfloat[]{\label{fig:etabarchart2.05}{\includegraphics[width=0.49\textwidth]{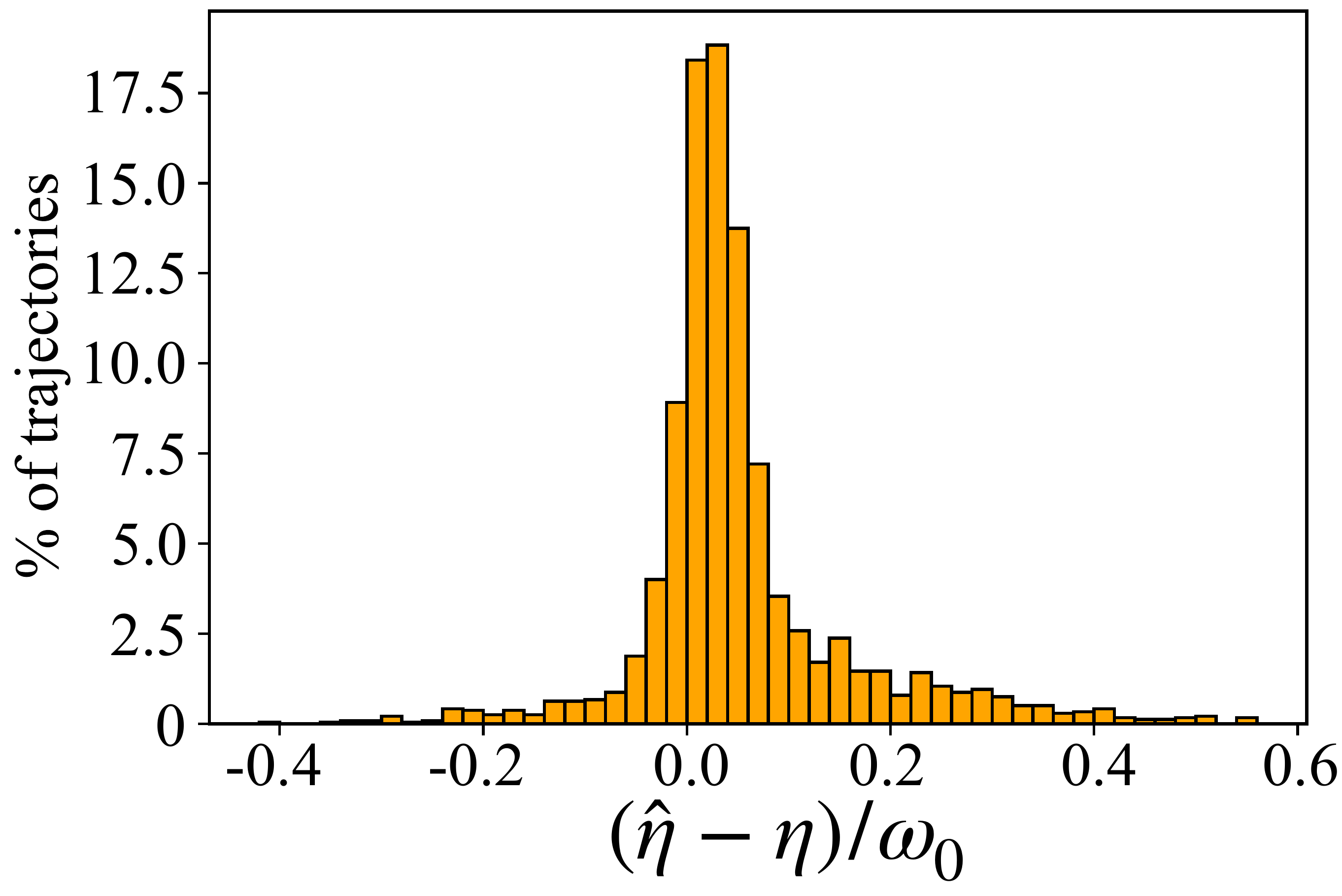}}}
\caption{Pure dephasing model: regression results for $\omega_c$, $s$, and $\eta$ for the largest interval length, where $\eta, \omega_c \in [0.25, 2.05]$. Panels (a), (c) and (e) display the predicted values $\hat{\omega}_c$, $\hat{s}$, and $\hat{\eta}$. respectively, plotted against the true values from the test set. A reference line is included in each plot to indicate ideal predictions, where the predicted values perfectly match the true values. Panels (b), (d) and (f) show bar charts depicting the error distribution, highlighting the percentage of trajectories whose prediction errors fall into specific interval for $\omega_c$, $s$, and $\eta$, respectively. }
\label{fig:regressionresults_0.25and2.05}
\end{figure*}

For the maximum interval length, where $\eta, \omega_c \in [0.25, 2.05]$, the NN faces more challenges when predicting $\eta$, $\omega_c$, and $s$ compared to the shortest interval. The MSE on the training set is $2.801 \times 10^{-3}$, while that on the test set is $3.049 \times 10^{-3}$. Although the small difference between these values indicates that the model generalises well and avoids overfitting, the higher MSE values reflect the complexity introduced by the broader range of parameter values. Figs.~\ref{fig:predictedomegacvreal_0.25and2.05},\ref{fig:predictedsvreal_0.25and2.05},\ref{fig:predictedetavreal0.25and2.05} show the predicted values against the true values for $\omega_c$, $s$, and $\eta$, respectively, for the trajectories in the test set. The corresponding error distributions for each parameter are shown in Figs.~\ref{fig:omegacbarchart2.05},\ref{fig:sbarchart2.05},\ref{fig:etabarchart2.05}. 

The plots of the predicted values against the true values clearly illustrate the increased difficulty the model faces as the interval length expands. For all three parameters, the model's performance is noticeably impacted by the broader parameter range, leading to greater deviations from the true values. In the case of $\omega_c$, the model's predictions remain reasonably accurate at lower values, but deviations become more pronounced at the higher end of the range. This suggests that the model has difficulty maintaining precision when $\omega_c$ approaches its upper bound. For $s$, the model exhibits larger errors at the lower end of the range, indicating that predicting sub-Ohmic values becomes particularly challenging as the parameter space widens. For $\eta$, the deviations are more spread out, with errors occurring for both intermediate and higher values. The error distribution for all parameters reveals a broader spread, with a noticeable increase in trajectories falling into larger error intervals. This broadening of the error distribution reflects the additional complexity introduced by the expanded parameter range, which makes it harder for the model to achieve the same level of precision as for the shorter interval. 

It is interesting to mention that qualitatively similar results can be found for non-Markovian amplitude damping dynamics: the corresponding numerical results can be found in the dedicated GitHub repository~\cite{github2}.

\section{Conclusions and Outlook}

We have explored a ML-based approach to the identification and characterization of spectral densities in open quantum systems. Focusing on an exactly solvable spin-boson model, we have shown that artificial NNs can be effectively trained to perform both classification and regression tasks of the relevant features of a SD, which brings about key information on the strength of the system-environment interaction. For both tasks, we used the temporal behavior of one of the observables of the system. 
On one hand, our classification results show that the network can accurately distinguish between sub-Ohmic, Ohmic, and super-Ohmic sSDs, achieving high accuracy even in challenging scenarios, where the coupling strength $\eta$ and the cut-off frequency $\omega_c$ are allowed to vary, alongside the Ohmicity parameter $s$. 
Similarly, the regression task -- targeting the estimation of $s, \omega_c, \eta$ -- yielded reliable predictions. 
The approach presented here can be extended to more complex scenarios involving structured SDs. For example, Ref.~\cite{Barr:2025} explores a related analysis in the context of open quantum dynamics where the spectral density is modeled as a sum of Lorentzian peaks.
These results support the applicability of ML as a valuable tool for learning large environments interacting with a quantum system, particularly when direct analytical or experimental access is limited.
The combination of these tools with experimental evidence could open new avenues for quantum noise identification and characterisation, as well as control of open quantum systems.
\\
\noindent
{\bf Acknowledgments} 
We acknowledge support from the European Union’s Horizon Europe EIC-Pathfinder
project QuCoM (101046973), the EIC Pathfinder Challenge project
Veriqub (101114899), the Department for the Economy of Northern Ireland under the US-Ireland R\&D Partnership Programme, the ``Italian National Quantum Science and Technology Institute (NQSTI)" (PE0000023) - SPOKE 2 through project ASpEQCt, the “National Centre for HPC, Big Data and Quantum Computing (HPC)” (CN00000013) – SPOKE 10 through project HyQELM.

%\bibliography{biblio}% common bib file


\end{document}